%% file: ms.tex
\documentclass[10pt,sigconf]{acmart}
\usepackage[utf8]{inputenc}
\usepackage{subcaption}
\usepackage[font=bf]{caption}
\usepackage{amsfonts}
\usepackage{booktabs}
\usepackage{amsmath}
\usepackage{color}
\usepackage{makecell}
\usepackage{bbding}
\usepackage{pifont}
\usepackage{wasysym}
\usepackage{amssymb}

\hypersetup{breaklinks=true}
\usepackage{algorithm}
\usepackage[noend]{algpseudocode}
\usepackage{balance}
\usepackage{enumitem}

\date{}

\usepackage{enumitem}
\newcommand{\squishlist}{\begin{itemize}[itemsep=1pt,parsep=2pt,topsep=3pt,partopsep=0pt,leftmargin=0em, itemindent=1em,labelwidth=1em,labelsep=0.5em]}
\newcommand{\squishend}{\end{itemize}}
\newcommand{\squishenum}{\begin{enumerate}[itemsep=1pt,parsep=2pt,topsep=3pt,partopsep=0pt,leftmargin=0em,listparindent=1.5em,labelwidth=1em,labelsep=0.5em]}
\newcommand{\squishsubenum}{\begin{enumerate}[itemsep=1pt,parsep=2pt,topsep=0pt,partopsep=0pt,leftmargin=0em,listparindent=1.5em,labelwidth=1em,labelsep=0.5em]}
\newcommand{\squishenumend}{\end{enumerate}}

\newcommand{\xref}[1]{\S\ref{#1}}
\renewcommand\footnotetextcopyrightpermission[1]{}
\newcommand{\name}{DeepSense}

\begin{document}
\title{DeepSense: Enabling Carrier Sense in Low-Power Wide Area Networks Using Deep Learning}

\author{Justin Chan$^{\ast}$, Anran Wang$^{\ast}$, Arvind Krishnamurthy, Shyamnath Gollakota\\
\normalsize{\normalfont{$^{\ast}$Co-primary Student Authors}}}

\maketitle

\input{abs-3}
\input{intro-7}
\input{case-4}
\input{design-3}

\input{spectrum-2}
\input{rnn-2}

\input{rate-2}
\input{eval-2}
\input{rate-eval-2}

\input{complexity-1}

\input{related-3}

\input{conc-1}

{\footnotesize \bibliographystyle{acm}}
\bibliography{noisynet} 

\end{document}

%% file: abs-3.tex
{\bf Abstract ---} The last few years have seen the proliferation of low-power wide area networks like LoRa, Sigfox and 802.11ah, each of which use a different and sometimes proprietary coding and modulation scheme, work below the noise floor and operate on the same frequency band.

We introduce DeepSense, which is the first carrier sense mechanism that enables random access and coexistence  for low-power wide area networks even when signals are below the noise floor.  Our key insight is that any communication protocol that operates below the noise floor has to use coding at the physical layer. We show that neural networks can be used as a general algorithmic framework that can learn the coding mechanisms being employed by such protocols to identify signals that are hidden within noise. Our evaluation shows that \name\ performs carrier sense across 26 different LPWAN protocols and configurations below the noise floor and can operate in the presence of frequency shifts as well as concurrent transmissions. Beyond carrier sense, we also show that \name\ can  support multi bit-rate LoRa networks by classifying between 21 different LoRa configurations and flexibly adapting bitrates based on signal strength. In a deployment of a multi-rate LoRa network, \name\  improves bit rate by 4x for nearby devices and provides a 1.7x increase in the number of locations that can connect to the campus-wide network.

%% file: intro-7.tex
\section{Introduction}
Low-power wide-area networks (LPWANs) have emerged as an enabling technology for connecting large numbers of sensors and devices across long ranges, at tens of milliwatts of power. These networks provide low-power connectivity between devices that are spread across many miles and enable various Internet-of-Things (IoT) applications including smart cities, utility management and asset tracking~\cite{apps,apps2}.

The IoT industry has made significant advances in this space over the last few years, with three critical trends emerging: Firstly, there has been a proliferation of numerous protocols supporting low-power wide-area networks including LoRa and Sigfox, with more being adding every year, e.g., NB-IoT, 802.11ah and others~\cite{nbiotprimer,halowm2m,zwave,dash7,weightless}. Secondly, each of these protocols differ in their throughput, range as well as their physical layer modulation and coding techniques. More importantly, many of these physical layer and link layer protocols including those used in LoRa and Sigfox are proprietary in nature~\cite{lorahome,sigfoxhome}. Finally, these protocols are being designed for the increasingly crowded 915~MHz ISM band and hence have to share the same set of frequencies since these competing technologies are simultaneously being deployed ---  Comcast is deploying its LoRa-based machineQ network across 12 major US cities~\cite{machineq}, while Sigfox has been deployed in over 100 US cities~\cite{sigfoxcities}.



In this paper we ask the following question: can we enable carrier sense for low-power wide area networks where devices use different protocols and configurations? A positive answer would enable multiple protocols to coexist in the 915~MHz band without interfering with each other; thus, allowing network operators to independently deploy these large-scale networks within the same metropolitan region.

\begin{table}[t]
\footnotesize
\centering

\begin{tabular}{| c  | c | c | c |}
\hline
{\bf Protocol} &  \thead{Data Rate\\(kbps)} & \thead{Bandwidth} & {\bf Modulation}  \\ \hline
LoRa SF 6  &5.85 -- 37.5&& \\ \cline{1-2}
LoRa SF 7  &3.41 -- 21.88&&    \\ \cline{1-2}
LoRa SF 8  &1.95 -- 12.50&&  CSS \\ \cline{1-2}
LoRa SF 9  &1.09 -- 7.03&125, 250, 500~kHz&{(Chirp Spread} \\ \cline{1-2}
LoRa SF 10 &0.61 -- 3.91&& {Spectrum)} \\ \cline{1-2}
LoRa SF 11 &0.34 -- 2.15&&  \\ \cline{1-2}
LoRa SF 12 &0.18 -- 1.17&&   \\\hline
LoRa FSK   & $<$ 1.2&2.6--250~kHz&FSK \\ \hline
Sigfox     & 0.1 -- 0.6&100~Hz&DBPSK and GFSK\\ \hline
NB-IoT     & $<$ 250&180~kHz&QPSK \\ \hline 
802.11ah   &150 -- 347,000&1,2,4,8,16~MHz&OFDM \\ \hline
\end{tabular}






\caption{\textmd{Popular LPWAN Protocols \& Configurations.}}
\label{table:sensitivities}
\vskip -0.25in
\end{table}

Achieving this goal however is challenging for multiple reasons: 1) Unlike traditional wireless networking technologies like Wi-Fi, it is difficult to use energy detection to perform carrier sense in these networks. Specifically, by the inherent nature of their long range operation, these technologies are designed to operate below the noise floor --- for example, LoRa can operate at SNRs as low as {-15~dB}~\cite{loramod}, where energy detection is difficult. 2) On the other hand, decoding a known preamble to perform carrier sense  under the noise floor does not scale with the growing number of protocols, each of which have multiple physical layer configurations. For instance, LoRa alone has 21 different physical layer configurations, each using a different physical layer code and producing a different preamble with a different carrier sense window size. Additionally, many of these protocols (e.g., Sigfox) are proprietary, and their exact physical layer coding and preamble structure may not be known. Thus, preamble detection below the noise floor, which requires understanding the underlying coding and modulation is difficult.  Finally, and importantly, such an approach is not forward-compatible since it requires modifying the hardware to add additional codes, modulations and preamble structures, every time a new protocol is introduced on the market.  As a result, today's LPWAN protocols  use either a centralized coordinator that does not support coexistence~\cite{caccess} or an inefficient ALOHA-based  MAC protocol~\cite{loralimits}. 

We introduce \name{}, which to the best of our knowledge, is the first carrier sense mechanism that enables random access and coexistence for low-power wide area networks.  Our design satisfies four key properties:

\squishlist
\item {\it Below-noise operation.} If a protocol is designed to operate below the noise floor, DeepSense can detect the corresponding signals even when the signal is below noise, without knowing the exact coding and modulation operation.

\item{\it Generalization.} It generalizes across various protocols including LoRa, Sigfox, NB-IoT* and 802.11ah as well as codes and modulations like FSK, QPSK, OFDM and chirp spread spectrum. It also does not require time or frequency synchronization information about the target protocols.

\item {\it Forward-compatible.} 
Our carrier sense design can work with new protocols without the need for upgrading the hardware. It can support carrier sense in the presence of future proprietary protocols by using a software update to update the weights used in our carrier sense algorithm.


\item {\it Scalability and Low-Power.} Finally, our carrier sense design has a computational complexity that is {\it independent} of the number of protocols, can operate in real time and more importantly work on low-power LPWAN radio hardware.

\squishend

Our key insight is as follows: any communication protocol that operates below the noise floor has to use coding at the physical layer to provide an SNR gain. Thus, a general algorithmic framework that learns the coding mechanisms employed by such protocols can be used to learn the codes and detect the presence of signals that are hidden within noise. 
By considering the coding operations employed by LPWAN protocols as continuous functions, we show in~\xref{sec:intuition} that, one can in theory use neural networks to perform carrier sense below noise.


Building on this intuition, we explore two deep learning  architectures that provide a tradeoff between power consumption, carrier sense window size, and training time. 
\squishlist
\item {\it Spectrogram+CNN.} The first approach is inspired by recent image de-noising systems~\cite{denoise} that can automatically restore the fidelity of images by removing noise using deep learning. To this end, we first compute a spectrogram over a fixed carrier sense window size. This effectively results in a compact and compressed real representation of the radio signals, which is similar to an image.  We  then train a single layer convolutional neural network (CNN) on this representation to identify LPWAN signals that are below the noise floor. 
\item {\it Dilated CNN+RNN.}  In the second approach, instead of using a fixed pre-determined carrier sense window, we utilize a dilated CNN architecture~\cite{wavenet2} to automatically learn a compressed representation of the wireless signals. Dilated CNNs are binary tree-like neural networks that have been used recently for compressing time series audio signals~\cite{wavenet2}. We train a recurrent neural network (RNN) on the output of this representation. 
This allows us to achieve a variable carrier sense window that dynamically accumulates probabilities to learn different window sizes for different protocol configurations as well as received signal strengths.
\squishend

We build a hardware platform consisting of a Raspberry Pi 3 CPU, which is connected via USB to a SDR~\cite{yoosoo}, and the Intel Movidius machine learning accelerator~\cite{movidius}. To create the training data, we capture over-the-air transmissions from a LoRa device that supports 21 different configurations, a LoRa FSK transmitter as well as Sigfox, NB-IoT*, RFID and 802.11ah transmitters in a {\it single location}. We then artificially simulate and introduce different wireless channel effects and noise to the training data set. Our test data is collected across eleven locations to span the whole operational SNRs for each of the tested protocols. This ensures that we are evaluating generalization across locations, over the air and different RF environments.
Our results show the following.

{
\squishlist
\item At an SNR of -10~dB, our carrier sense system can detect a LoRa SF11 signal with an accuracy of 99\%.
We can also perform carrier sense across LoRa, FSK, Sigfox, RFID, NB-IoT* and 802.11ah and can provide accuracies up to 97\%. Further, using a fixed window size, we can detect RFID signals at SNRs that were better than what they were designed for. 
\item The RNN approach can provide variable carrier sense windows {at 0.8~ms increments}. Further it achieves an accuracy of 88\% for SNRs at -10 to -15~dB. This is higher than the spectrogram approach which achieves an accuracy of 61\%. This is partly because the RNN preserves phase information that is discarded by the spectrogram operation.
\item While the training data only had frequency shifts of up to 10~Hz, our system can detect signals that are offset by frequency shifts as high as 250~kHz. It can also perform carrier sense with concurrent transmissions within the receiver's bandwidth from the same as well as different protocols.
\item For continuous carrier sense operations, the two architectures  require {348M} and 849M FLOPS. We estimate the power consumption on a deep learning ASIC  as 9--11~mW. For comparison, a LPWAN radio consumes 30--50~mW~\cite{power}.
\squishend
}

Beyond carrier sense, the ability to identify protocol configurations from signals that are significantly below the noise floor can also enable LPWANs where devices use different bit rates depending on their location. Specifically, although LoRa supports 21 different physical layer configurations and preambles, current LoRa networks use a fixed bit rate since the receivers can only recognize a preset preamble configuration. We show that a \name{}-enabled LoRa receiver can classify between all the 21 LoRa physical layer preambles at low-power and hence can support a multi bit-rate LoRa deployment. To this end, we deploy a multi-rate LoRa network in a large campus area that flexibly adapts the bitrate based on the signal strength (see~\xref{sec:rateeval}). Our results show that \name{} can classify between LoRa configurations at their designed sensitivities, with an average accuracy of 95\% for SNRs at -10 to 5~dB. Further, compared to a single-rate network operating at 9.38~kbps, our multi-rate LoRa network can support all the LoRa bit rates from 183~bps to 37.5~kbps resulting in bit rate improvements of 4x for nearby devices and a 1.7x increase in the number of locations that can connect to the network.

\vskip 0.05in\noindent{\bf Contributions.} We present the first carrier sense mechanism that enables random access and coexistence for LPWANs. To do this, we first present a theoretical analysis that shows that neural networks can be used for learning various codes in LPWAN networks. We then present two different deep learning architectures to achieve real-time and low-power carrier sense capabilities that work with protocols that operate below the noise floor. Finally, we show that this approach can be used to classify between different LoRa configurations and enable multi-rate LPWAN networks.

%% file: case-4.tex
\section{Understanding the problem}
In this section, we first provide an overview of recent developments in the LPWAN space. We then motivate the need for carrier sense in these networks.

\subsection{Overview of LPWAN Protocols}
\label{sec:lpwan}

Table.~\ref{table:sensitivities}  lists various LPWAN protocols that have seen some adoption in the industry in recent years. The table includes a number of these popular protocols and standards such as LoRaWAN and 802.11ah have been introduced recently between 2015~\cite{lorawanout,halowout} and 2017, showing that this is a fast evolving industry. These protocols, including 802.11ah which is a recently published standard designed by the Wi-Fi alliance, are designed to operate at 915~MHz.

The table highlights various important features of these protocols that are important for our design.
\squishlist
\item  Many of the physical and link layer properties of these protocols (e.g., LoRa and Sigfox) are proprietary in nature. 
\item These protocols use different bandwidths, achieve different bit rates and their preambles occupy different durations on the channel. They also use a range of modulation and coding techniques including {FSK (frequency shift keying), QPSK (quadrature phase shift keying), OFDM (orthogonal frequency-division multiplexing) as well as CSS (chirp spread spectrum).}  This make it hard to inter-operate between devices talking these different protocols.
\item Many of these protocols have a number of different configurations: LoRa networks can be configured with three different bandwidths (125, 250 and 500~kHz) and seven different spreading factors (SFs) resulting in a range of bit rates from 183~bps to 37.5~kbps. Since each of these 21 configurations results in a  different preamble structure, existing LoRa networks are preset to use a single configuration.
\item These protocols can be decoded at low powers well below the noise floor and hence have a long range. LoRa configurations can be received at sensitivities as low as -137~dBm (-15~dB SNR)~\cite{loramod} while Sigfox can be decoded at a sensitivity of -126~dBm (-4~dB SNR). In contrast Wi-Fi signals require a signal strength more than -90~dBm to be decoded.
\squishend

These protocols achieve such below-noise operation by using coding. For instance, LoRa uses a physical layer code called chirp spread spectrum (CSS) where data is encoded using upchirps where the frequency of the signal linearly increases in time. The receiver achieves a coding gain by multiplying this signal with a downchirp where the signal frequency linearly decreases with time allowing for below-noise operation. This code however is different across different protocols (e.g., chirps are only used in LoRa) and is also different for various configurations of LoRa, i.e., different bandwidths and spreading factors use a downchirp with the corresponding bandwidth and spreading factor as the code.

\subsection{Case for Carrier Sense}

Since different LPWAN protocols operate under the noise floor and cannot decode each other, currently they use ALOHA based random access where nodes simply transmit packets when they have data. While it is well known that such an ALOHA-based MAC protocol has an efficiency of 18\%~\cite{apps2}, the problem is made severe since wide-area city-wide networks can have a large number of devices. Moreover, since these protocols use low bit rates, they can occupy the medium for a long time, increasing the probability of collisions. For example, the longest LoRa packet when using a payload of 50 bytes is over 3 seconds on the wireless medium (when the bandwidth is 125~kHz and spreading factor is 12).
Sigfox packets for the same payload are over 1.5 seconds. For comparison, Wi-Fi transmissions occupy around a millisecond.

Due to the long packet lengths and an ALOHA based random access, LPWAN networks have a high probability of collisions. To understand this, we consider two cases.

\begin{figure}[t!]
\begin{subfigure}[t]{0.23\textwidth}
\includegraphics[width=\textwidth]{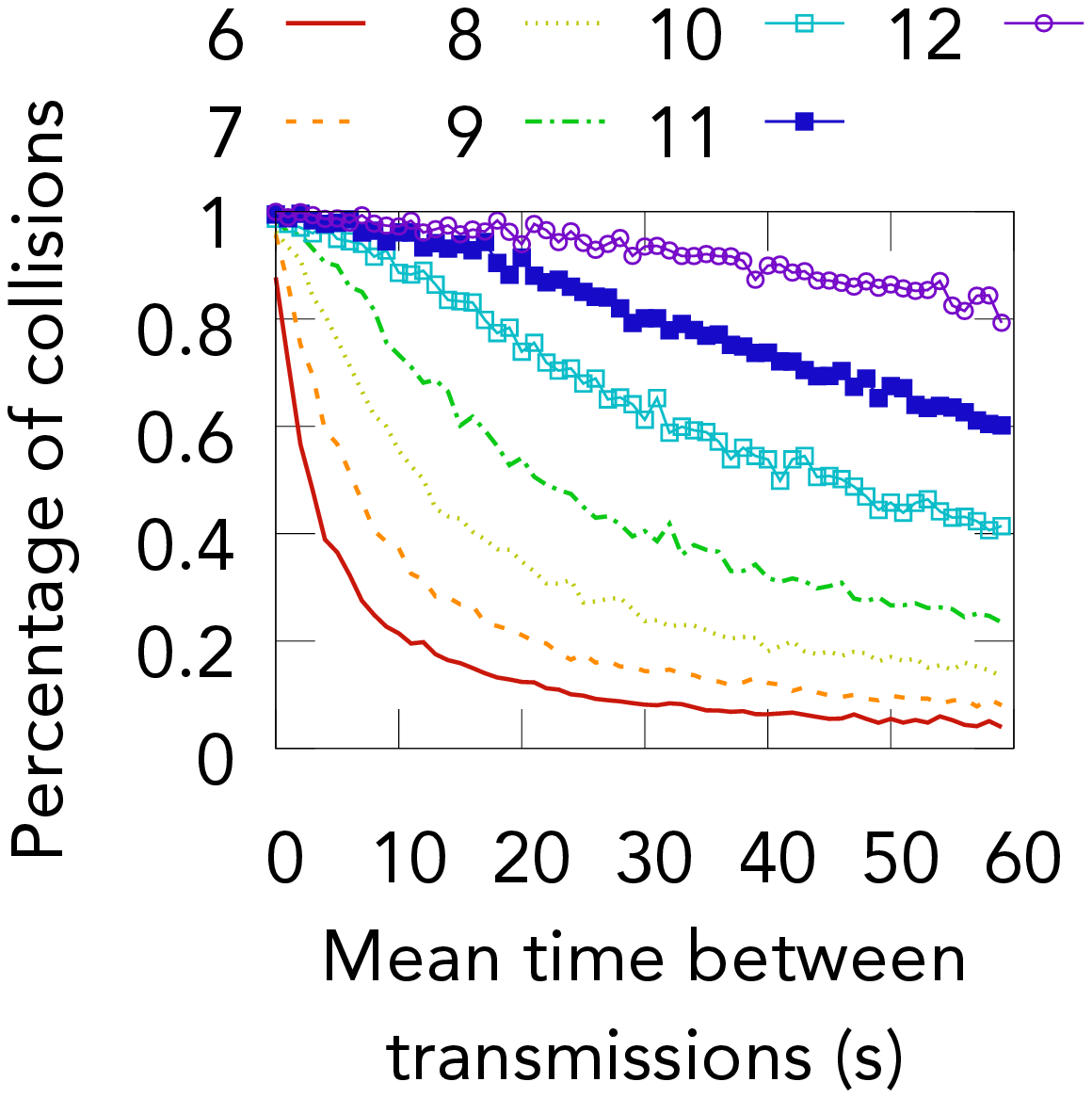}
\caption{\textmd{100 nodes}}
\end{subfigure}
\begin{subfigure}[t]{0.23\textwidth}
\includegraphics[width=\textwidth]{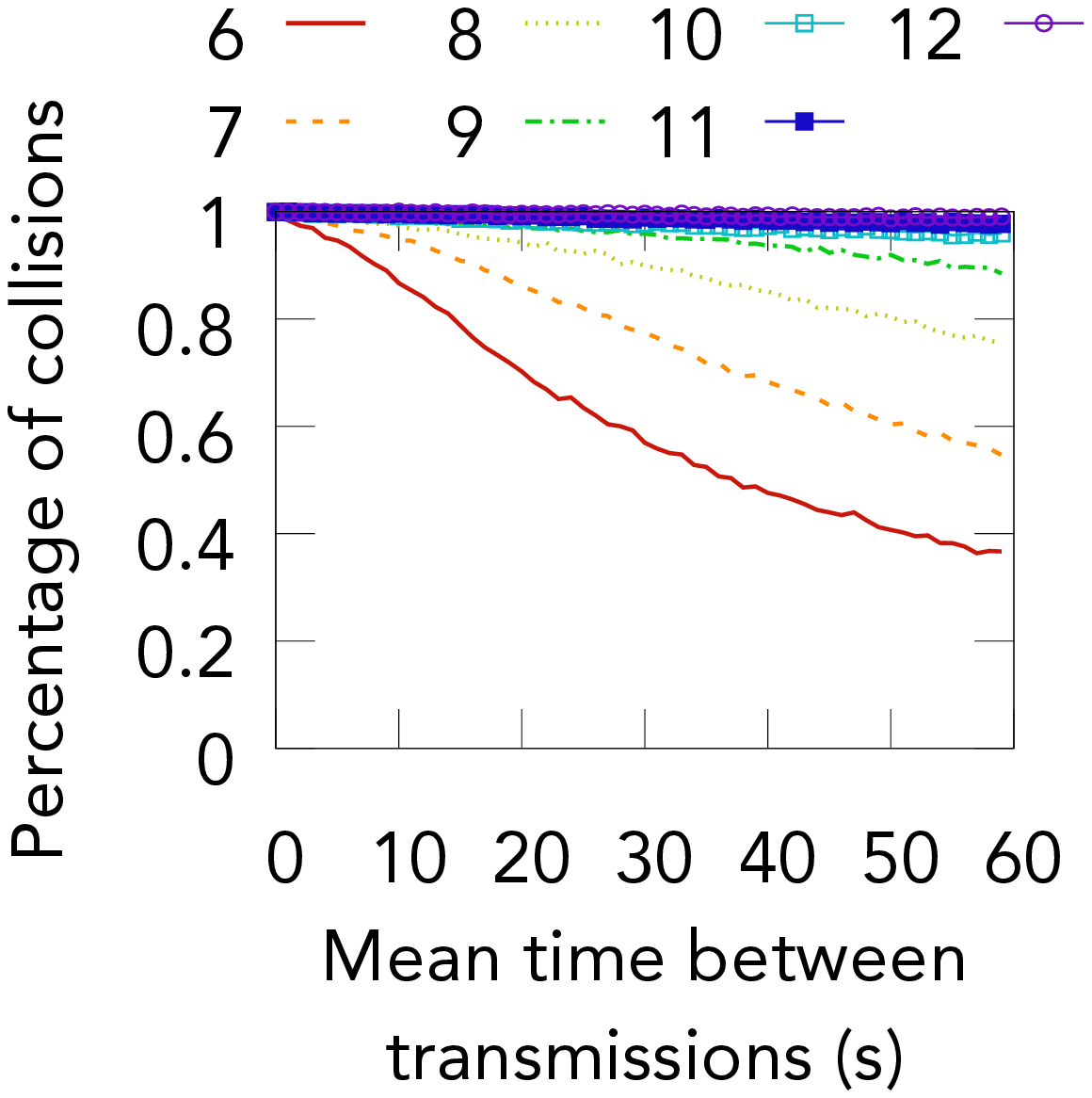}
\caption{\textmd{1000 nodes}}
\end{subfigure}
 \vspace{-0.1in}
\caption{\textmd{Percentage of collisions in a fixed configuration LoRa network across different spreading factors.}}
 \vspace{-0.2in}
\label{fig:single}
\end{figure}

{\it Collisions within a single LPWAN protocol.} Consider a $N$ node LoRa network across a metropolitan city, where each node transmits 25-byte packets periodically and where the periodicity follows the exponential distribution with mean $M$. Fig.~\ref{fig:single} shows the probability of collisions when a 100-node and 1000-node LoRa network is run at different spreading factors using a 250~KHz bandwidth as a function of $M$ over the course of an hour. We use the LoRaSim~\cite{lorasim} tool to compute these results. The plots show that the collision probability is reasonably high. Additionally, networks with higher spreading factors and more devices result in more collisions. This leads to high packet losses in LoRa networks with large number of devices~\cite{loss1,loss2,loss3}. This has led to recent work on decoding collisions in LoRa networks~\cite{lora-sigcomm17}.

\begin{figure}[t!]
\begin{subfigure}[t]{0.23\textwidth}
\includegraphics[width=\textwidth]{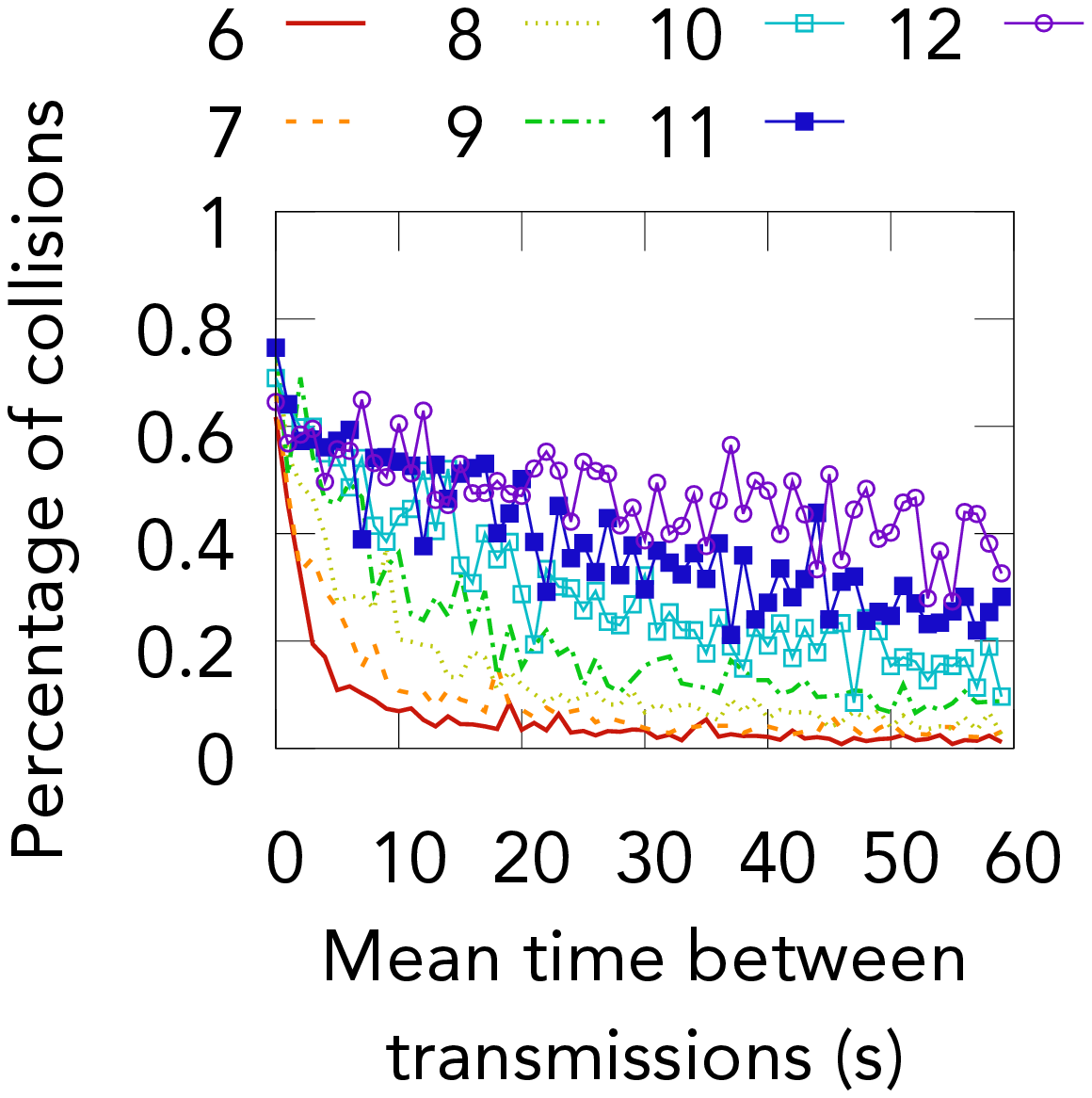}
\caption{\textmd{100 nodes}}
\end{subfigure}
\begin{subfigure}[t]{0.23\textwidth}
\includegraphics[width=\textwidth]{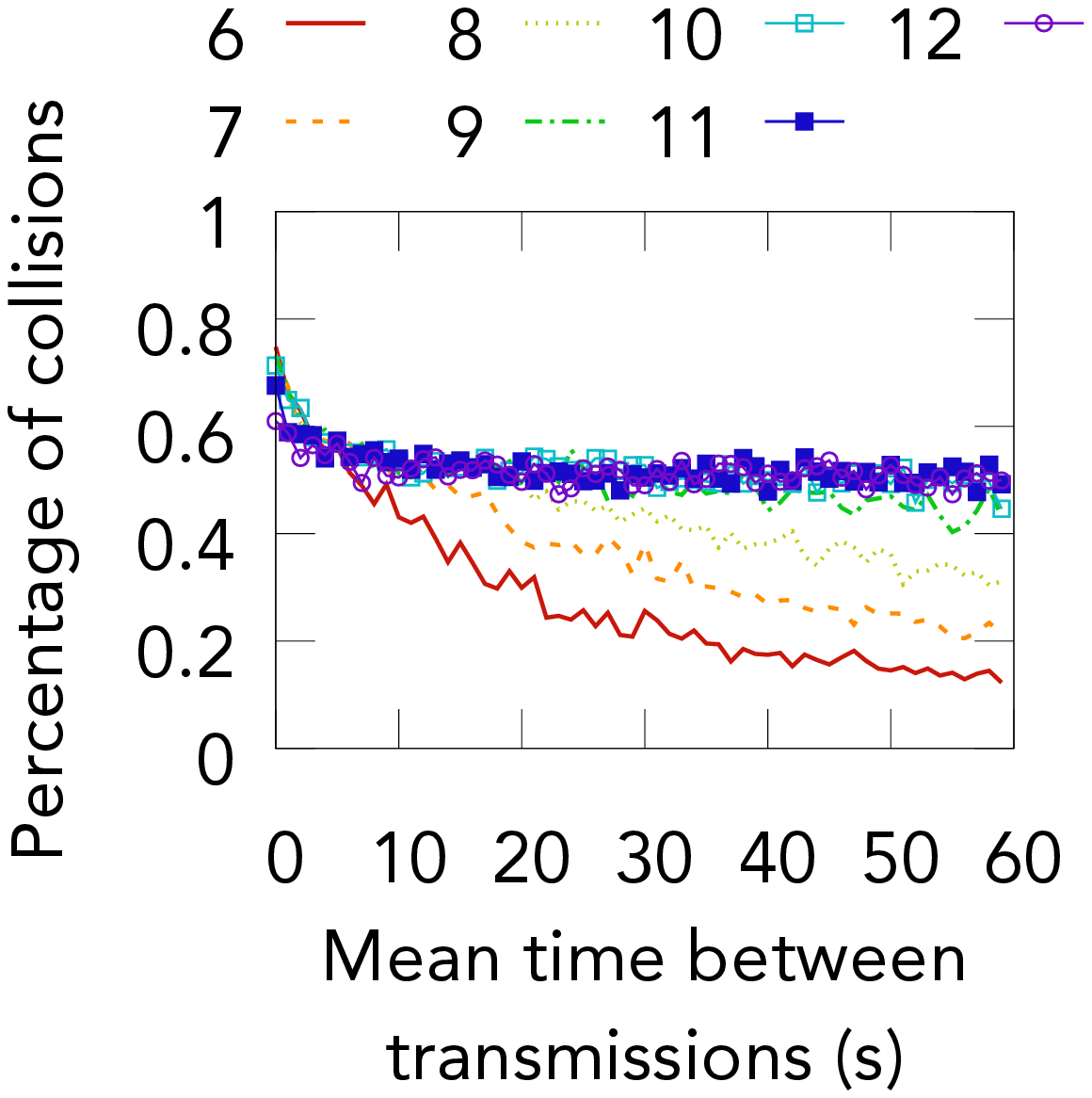}
\caption{\textmd{1000 nodes}}
\end{subfigure}
 \vspace{-0.1in}
\caption{\textmd{Percentage of collisions in a LoRa and Sigfox network across different LoRa spreading factors.}}
\vspace{-0.15in}
\label{fig:lorasigfox}
\end{figure}

{\it Collisions across LPWAN protocols.} Next, let us consider two co-located LPWAN networks in the same metropolitan area run by two different operators (e.g., T-mobile and Comcast). One operator is running a LoRa network, and the other is operating a Sigfox network. We examine the case when the total number of nodes across the networks is 100 and 1000, with each network having an equal number of nodes. As above, the LoRa network transmits 25-byte packets. The Sigfox nodes transmit a 24-byte packet with a 12-byte payload {at 100~bps} over a duration of 1.92~s. The Sigfox packets have a bandwidth of 100~Hz and use frequency hopping. Fig.~\ref{fig:lorasigfox} shows the probability of collisions in the above deployment for different LoRa spreading factors. 



The above empirical results show that, as expected, ALOHA based systems result in a significant number of collisions in dense deployments that are typical of city-wide networks. This motivates the need for a carrier sense based solution that can operate across different LPWAN protocols. 

%% file: design-3.tex
\section{Carrier Sense with \name}
\label{sec:intuition}
Designing an accurate and efficient carrier sense that works across LPWAN protocols is challenging for  three reasons:
\squishlist
\item Our system should  work across multiple protocols and coding schemes. It should also be forward-compatible and support the addition of new protocols  without hardware changes or decreases in computational efficiency.
\item Carrier sense requires real-time operation. As such our system should be able to distinguish a signal from noise in less time than it would take for that signal to be transmitted. 
\item Our design should operate on complex I-Q signals that are sampled at say 1~MS/s with a {16--bit} resolution for each of the I and Q samples. Processing this amount of data with a low delay, requires compressing the incoming wireless data.
 
\squishend


Our intuition in designing a carrier sense mechanism is to leverage the Universal Approximation Theorem for feed-forward neural networks~\cite{cybenko}:

\begin{theorem}
Let ~$C(I_m)$ be the space of non-constant, bound -ed 
 and monotonically-increasing continuous functions on an m-dimensional unit hypercube $I_m$. {For any $\epsilon>0$,} a feed forward neural network with at least one hidden layer of a finite number of units followed by a non-linear continuous activation function can produce a function $F(x)$ that approximates any $f \in C(I_m)$ such that $|F(x)-f(x)| < \epsilon$ for all $x \in I_m$, where  $I_m$ can also be any compact subset of  $\mathbb{R}^n$.
\end{theorem}

At a high level, {neural networks achieve this by reducing the difference between $F(x)$ and $f(x)$, for a given $x$}. A feed-forward neural network first passes its inputs through a set of weights in the network to generate an initial estimate of $f(x)$. It then calculates the error between $f(x)$ and $F(x)$ using a loss function. Backpropagation is then used to calculate the gradient of the loss function using these error values. {Based on these gradients, an optimization method is used to iteratively reduce the loss value using a new set of weights.}

\vskip 0.05in\noindent{\bf Deep learning for carrier sense.} Wireless signals can be represented as a stream of complex numbers, $\mathbf{x}$, where $\mathbf{x[n]}$ is the $n^{th}$ transmitted sample. The received signal $\mathbf{y}$ of a narrowband flat-fading channel can be approximated as \\ $\mathbf{y[n] = Hx[n]+w[n]}$, where $\mathbf{H}=he^{\gamma}$ is the complex channel, $h$ represents signal attenuation over distance and $\gamma$ refers to the phase shift between the transmitter and receiver. $\mathbf{w[n]}$ represents additive white Gaussian noise.
A wideband channel with multipath can be approximated by the summation of $L$ different multipaths:
$
\mathbf{y[n] = \sum_{i=1}^L H_ix_{i}[n]+w[n]}
$.

So, carrier sense over a set of protocols can be defined as, 
\begin{definition}
A generalized carrier sense over any set of protocols $P$ can be modeled as a function $F(\mathbf{y})=\bigvee_{p\in P} F_p(y) \in \{0,1\}$, where $\mathbf{y}$ is the wireless signal. 
\end{definition}

Unlike audio and video signals, wireless signals are represented as complex numbers and are not in $C(I_m)$ or $\mathbb{R}^n$. One challenge with using complex numbers with neural networks is that a complex valued function $f(x)$ that is differentiable will be at least unbounded~\cite{complexcomp}. As such when the function $f(x)$ is passed through a nonlinear activation function like $tanh$, it will have singularity points which go off into infinity. As a result, the neural network may not converge to a good approximation $F(x)$~\cite{complexcomp}. While recent works~\cite{iclrcomplex,complex2} have built complex-valued neural networks by creating new activation functions and operations for complex numbers, they rely on specialized architectures which are difficult {to generalize to any set of inputs}.

To address this problem, our design first passes the wireless signals through a transform function:

\begin{equation*}
\mathbf{z = transform(y):\mathbb{C}^n \mapsto} \mathbb{R}^{m \times k}
\end{equation*}

Said differently, the transform function transforms the $n$ complex samples into $k$ channels of $m$ real samples. {For a spectrogram, $k$ and $m$ refer to frequency and timing information respectively.} After this transformation, the following lemma follows from the universal approximation theorem.  

\begin{lemma} \label{eq:lemma}
We define the carrier sense function on the real domain as $F_R:\mathbb{R}^{m \times k}\mapsto [0,1]$. \\
If for every $x\in \mathbb{C}^n$, $|F_R(\mathbf{transform(x)})-F(x)|<\delta$ for some $\delta>0$, 
there exist a neural network $f_R(y)$ that can approximate $F_R(\mathbf{transform(x)})$ such that \\
$|f_R(\mathbf{transform(x)}) - F(x)| < \epsilon+\delta$ for some $\epsilon > 0$.
\end{lemma}

Based on this, we posit that given a good transform function and enough data, deep neural networks should be able to learn a carrier sense mechanism for any set of protocols. In communication systems, exponential amounts of learning data can be automatically generated  by  changing the bits. Thus, by learning carrier sense, our system would be able to detect the presence or absence of a packet even when it is below the noise floor, and thus provide carrier sense capabilities for LPWAN protocols.

In addition to the correctness property discussed above, neural networks also meet our four design criteria:
1) As neural networks are universal function approximators, they would in theory be able to approximate all LPWAN codes. This also allows \textit{below-noise operation}, 
2) Neural networks can learn different LPWAN codes using the same architecture, and thus support \textit{generalization}, 
3) Neural networks are \textit{forward-compatible}. By updating their weights with a software update, neural networks can learn new codes and support future proprietary protocols,
4) Finally, using machine learning accelerator ASICs~\cite{power1, power2, power3}, 
neural networks can make inferences at a \textit{low power}. 


Building on the above theory, we present two complementary deep learning architectures that enable carrier sense under the noise floor. Each architecture has three parts. A transform function that maps complex wireless signals to real numbers, and allows neural networks to approximate a carrier sense function. A compression function that reduces training and inference times, and enables a low-power carrier sense scheme. And finally a classification function which maps the input representations to either signal or noise.


%% file: spectrum-2.tex
\subsection{Spectrogram+CNN Architecture}

Our first architecture is inspired by image de-noising systems~\cite{denoise} that use deep learning to automatically restore the fidelity of images that are impaired by noise. 

\vskip 0.05in\noindent{\bf Spectrogram as the transform and compression.} In this approach, we transform a fixed window of complex IQ samples into real values using a spectrogram. The spectrogram preserves timing, frequency and power information about the signal in the form of a two-dimensional array of power values, {where the $x$ axis represents frequency and the $y$ axis represents the time}, that is similar to an image. Our main intuition behind this approach is that modulation schemes like CSS, FSK, PSK and OFDM are continuous over frequency and time domains. As such, information is spatially related and pixels within a local region are more closely related than pixels that are further away. This transform process places our signals in $C(I_m)$, and by Lemma \ref{eq:lemma} a neural network would be able to approximate the optimal carrier sense scheme.

More formally, the spectrogram is first computed by taking the short-time Fourier transform on complex IQ samples, $STFT(\mathbf{x(n)},m,\omega) = \sum_{n=-\infty}^{\infty} \mathbf{x(n)} wind[n-m] e^{-j\omega n}$, where $wind$ is the Hann window and $m$ is the window size. To get the spectrogram we compute the power of this short-time Fourier transform function, $|STFT(x,m,\omega)|^2$. 
This operation also compresses signal inputs with {window} size $B$ into a 2D array of $b$ real numbers where $B\geq b$. 

Our implementation takes the spectrogram over a fixed window size of up to 8~ms and uses a 64-point discrete Fourier transform, resulting in a spectrogram spanning 64 by 39 values. With a {window} size of 8000 complex samples, the spectrogram compresses the samples by {84\%} into {2496} real  values. 

\begin{figure*}[t!]
\vspace{-0.15in}
\includegraphics[width=0.18\textwidth]{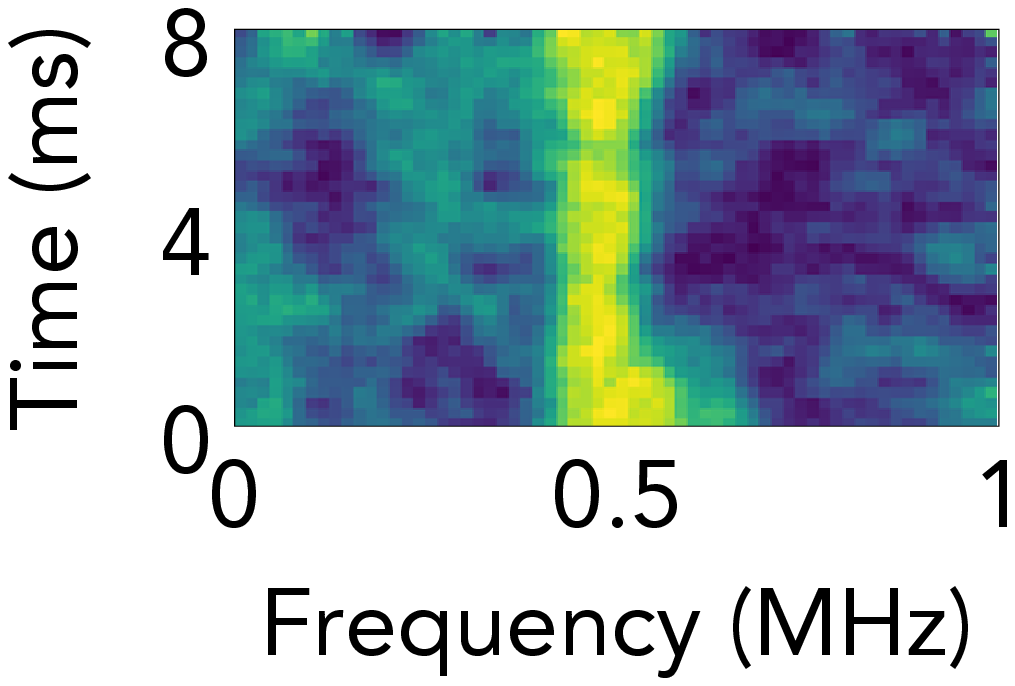}
\includegraphics[width=0.18\textwidth]{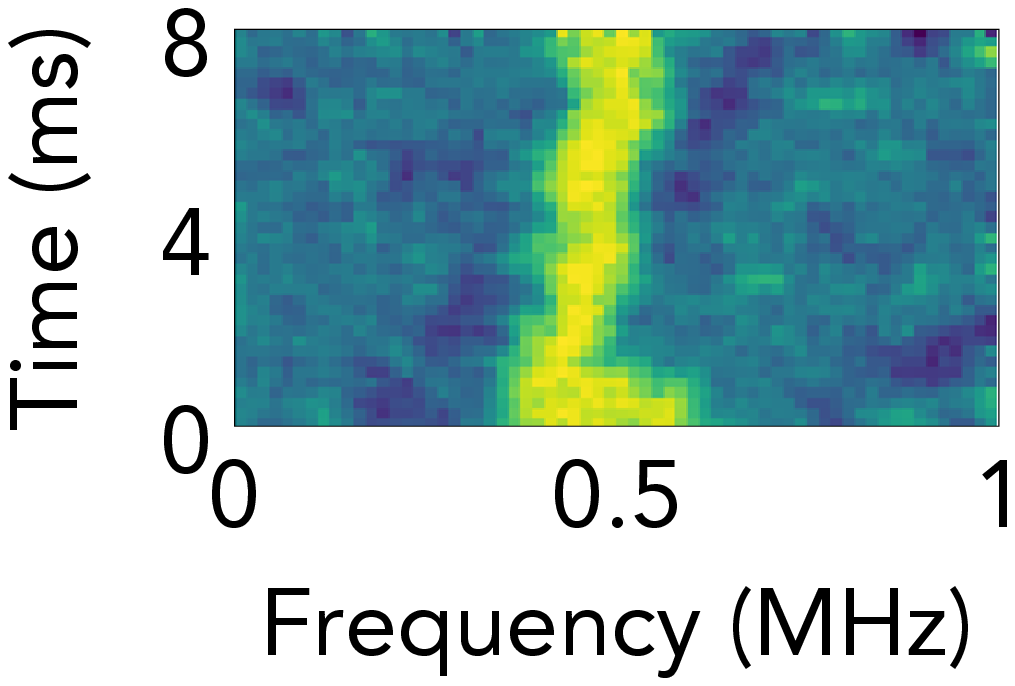}
\includegraphics[width=0.18\textwidth]{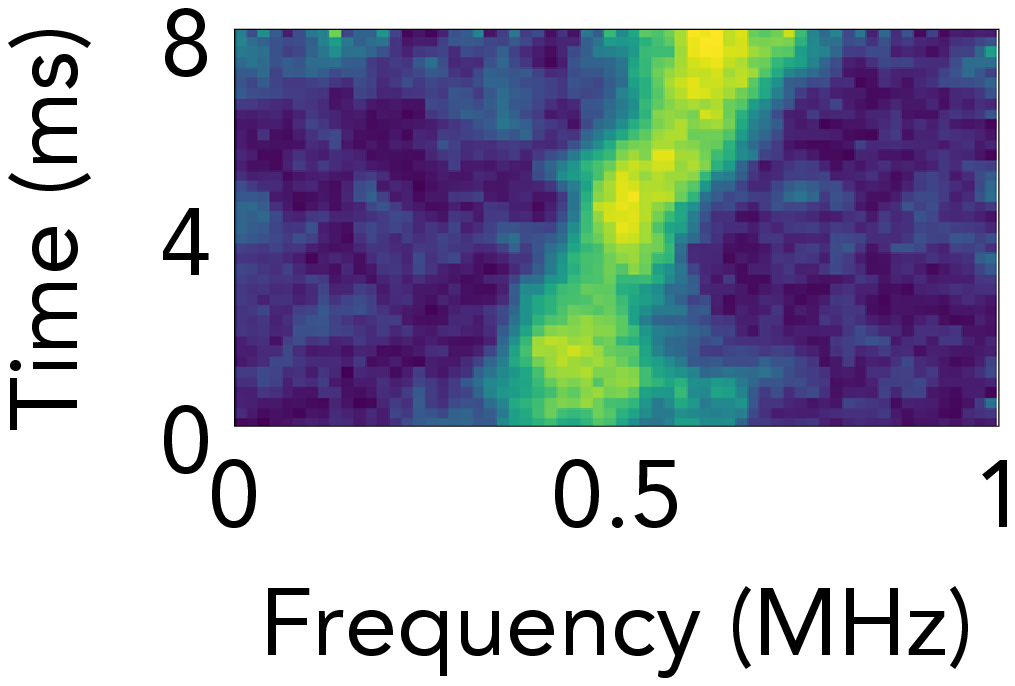}
\includegraphics[width=0.18\textwidth]{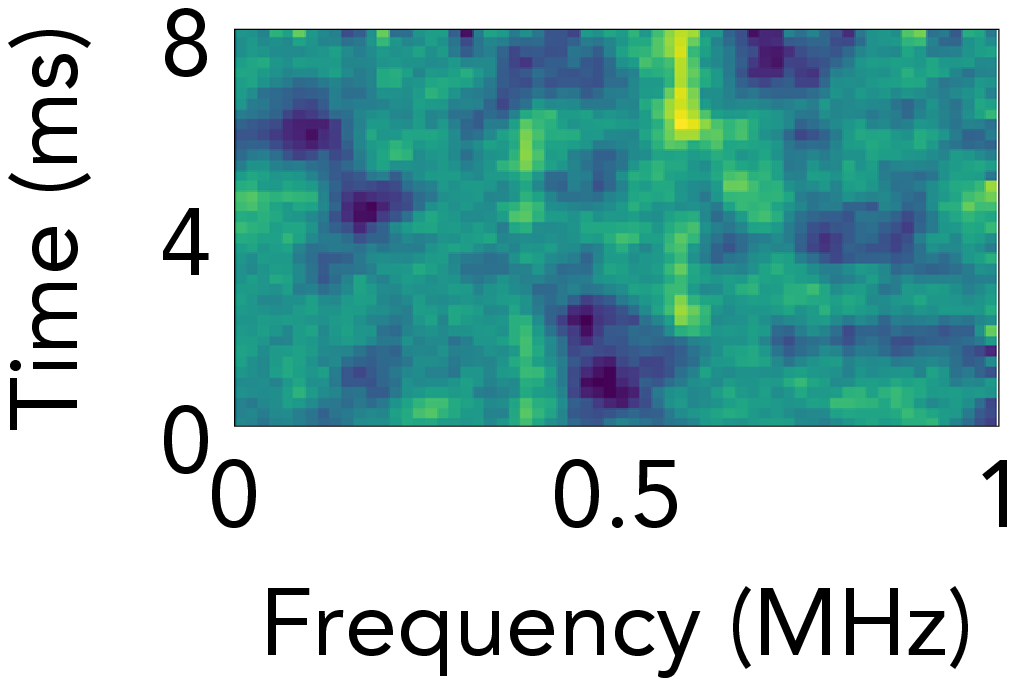}
\includegraphics[width=0.18\textwidth]{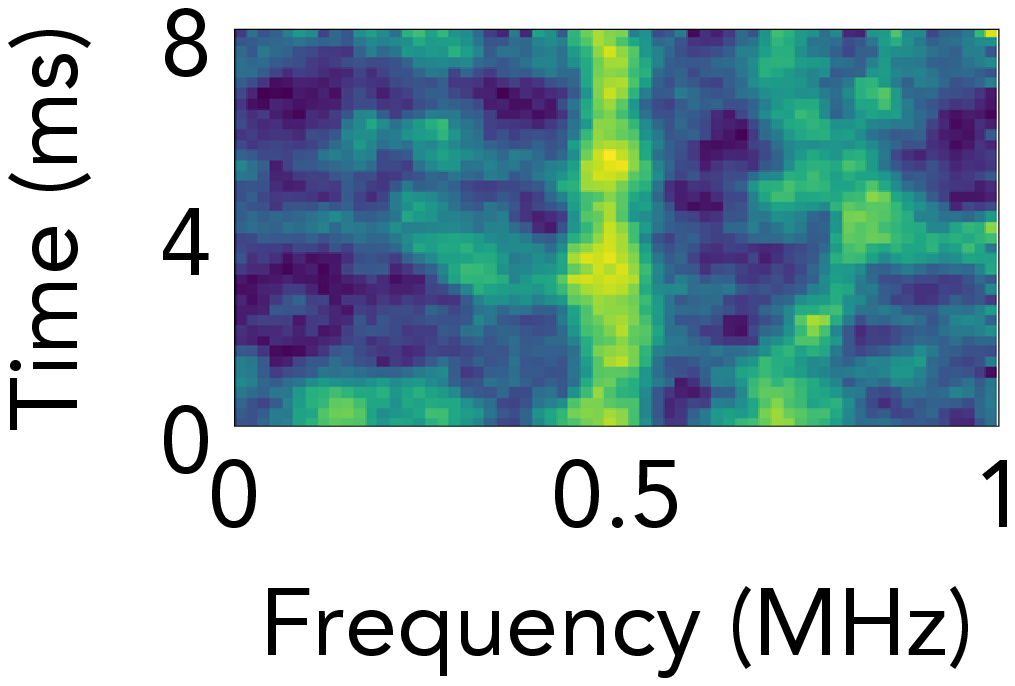}
\vskip -0.15in
\caption{\textmd{Visualizations of learned signals. a) LoRa - 125~kHz bandwidth b) 250~kHz c) 500~kHz d) FSK e) Sigfox.}}
\vskip -0.2in
\label{fig:viz}
\end{figure*}

\vskip 0.05in\noindent{\textbf{Convolutional Neural Networks as classifiers.}}
Convolutional Neural Networks (CNNs) are a natural network architecture to use when training over images and spatially-correlated signals (e.g., spectrograms), which are correlated in both time and frequency. CNNs are also robust to translations of the input data. As a result our architecture can work with frequency-and time-offsets that are typical in practical wireless signals. CNNs also work well with noise as the convolution operation creates smoothed internal representations on the input spectrograms and act as a low-pass filter over noise. Note the similarity with the convolution operation used in wireless systems to decode signals below the noise floor by looking for a known signal pattern.


When the spectrogram of dimensions $W$ and $L$ passes through the convolution layer, an $N$ by $N$ kernel ($N=3$) is convolved at every point in the image in a sliding window fashion. This produces an $W-N+1$ by $L-N+1$ image. This process is repeated $F=64$ times with different kernels to produce $F$ filters. Each kernel is initialized with zero mean and unit variance and the kernel values are learned over time with backpropagation. Each filter is then passed into the ReLu non-linear function, $\sigma(f) = max(0,f)$.

The filters are then passed to an average pooling layer which reduces the number of parameters in our model and prevents overfitting. These outputs are vectorized into a single column.
Since carrier sense requires distinguishing between two classes, the final layer of our network is a fully connected layer with two units. The column vector is multiplied and summed by the weights and biases in the final layer, then passed to a \textit{softmax} function to output a probability indicating which class the input spectrogram belongs to. {After multiple iterations of back-propagation and stochastic gradient descent, the network converges onto a desired set of filters, weights and biases. 

\vskip 0.05in\noindent{\it Visualizing the learned signals.}  To better understand the above process, we visualize the wireless codes that are learned after the CNN is trained on a given dataset. {To do this, we first transform the raw set of LoRa chirps into spectrograms. We implement the above design and} train our model on 1000 spectrograms of LoRa chirps with a spreading factor of 10,  bandwidth of 125~kHz and 10~dB SNR, and 1000 spectrograms of artificially generated noise. After training the network, we use~\cite{kerasviz} to generate a spectrogram image that maximizes the probability the network will classify the image as a chirp. We repeat this process for LoRa chirps with bandwidths of 250~kHz and 500~kHz, a LoRa FSK signal and a Sigfox DBPSK signal. Fig.~\ref{fig:viz} shows the learned spectrograms for all these signals. The visualizations for the LoRa chirps show that each chirp occupies a different bandwidth. The FSK spectrogram shows the symbols being modulated between two frequencies. And the Sigfox spectrogram shows a narrowband 200~Hz signal. These signals appear to occupy a larger bandwidth as the smallest FFT bin size in our spectrogram is 15~kHz. We note that our architecture and hyperparameter settings were the same when training all these protocol-specific carrier sense schemes. This shows that our technique is general enough that it can be applied to learn the structures of multiple signals.

%% file: rnn-2.tex
\subsection{Dilated CNN + RNN Architecture}

The above approach is limited as the spectrogram  operates on a fixed window size which cannot be adapted at runtime. While we can set the window size to be the same as the shortest preamble length, applying the same window size to protocols with much longer preambles may result in loss of information and lower accuracies. {Moreover, the spectrogram representation discards phase information, which is important when characterizing between phase-based modulations like BPSK and QPSK.} 

\begin{figure}
\hspace{-0.1in}
\includegraphics[width=0.48\textwidth]{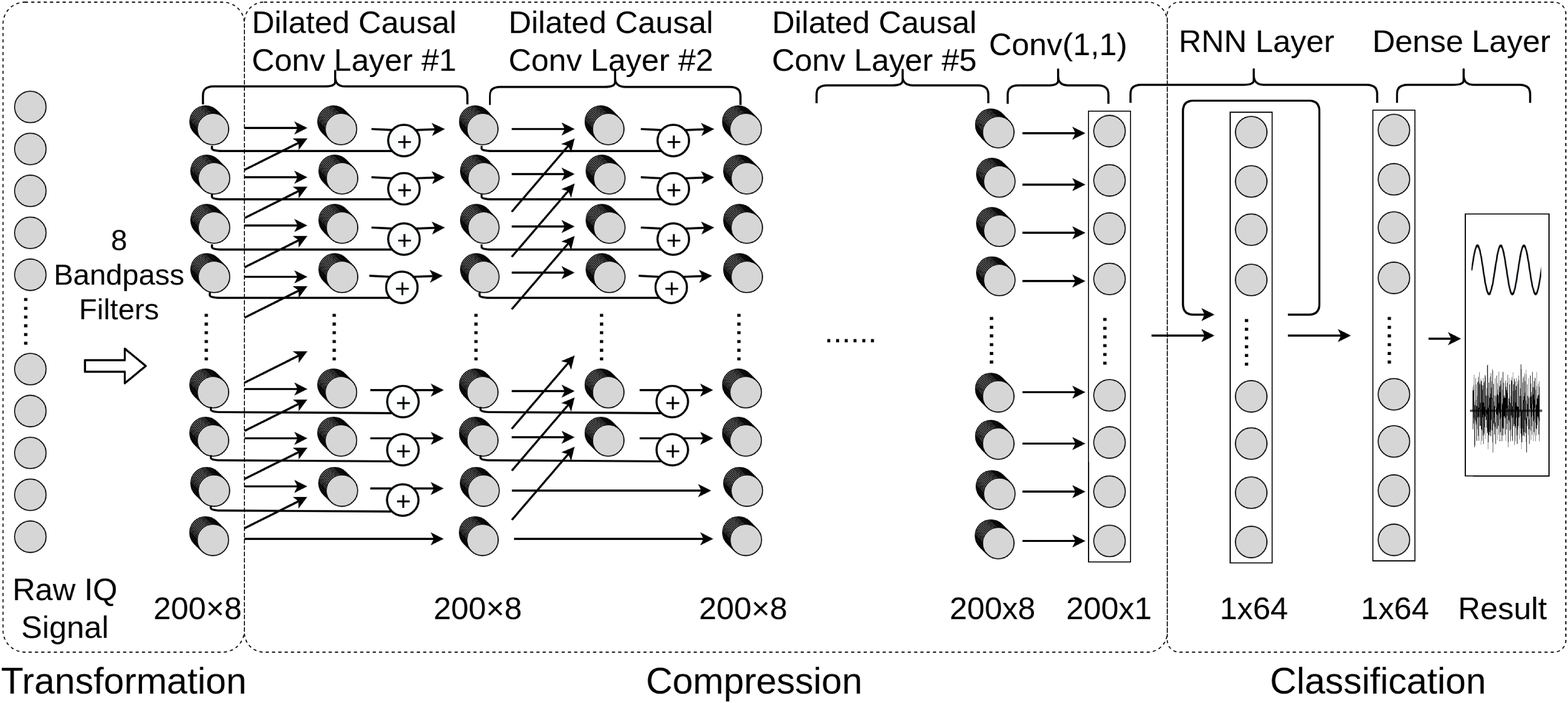}
\vskip -0.1in
\caption{\textmd{Dilated CNN + RNN architecture.}}
\vskip -0.2in
\label{fig:wavenet_rnn}
\end{figure}

Our ideal design should adaptively choose the length of the carrier sense {window} for each protocol. To have an adaptive {window} size, our design should support finer-grained input units, and accumulate information after processing each unit. To this end, our second approach in Fig.~\ref{fig:wavenet_rnn} uses a recurrent neural network (RNN) that provides the above capability.

\vskip 0.05in\noindent{\bf Sub-band splitting as the transform function.} Similar to the first approach, we first need to transform the raw complex IQ samples into a real representation suitable for neural networks. We use eight band-pass filters to split the 1~MHz band into eight 125~kHz subbands. We map each subband into the real frequency range from 0 to 125~kHz. We then sample each subband at 250~kHz to convert our complex samples into real samples. We use 800 complex samples as input (corresponding to 0.8~ms). After the above transformation, we get a $200\times 8$ real sample matrix for each block. The intuition for this transform is that sub-band splitting divides the spectrum into bands which contain useful signals, and bands which just contain noise. {Note that we lose neither phase nor amplitude information after this transformation.}

\vskip 0.05in\noindent{\bf Dilated causal convolutions as compression.} Next, we let a neural network compress the above data into a compact format.  {To achieve this, we use a technique known as dilated causal convolutions that was used in Google's WaveNet project~\cite{wavenet2} to achieve state of the art speech synthesis.}

As seen in our architecture in Fig.~\ref{fig:wavenet_rnn}, our input layer has 200 units, where each unit has eight values. This is passed through a sequence of $L=5$ dilated causal convolutional layers. For each unit $i$ in layer $l$, {we first calculate $\mathbf{y}_{l,i}$ as a linear combination of the output of the two units in the previous layers weighted by learned weights $W_1$ and $W_2$:}
$$ \mathbf{y}_{l,i}=\mathbf{z}_{l-1, i-2^{l-1}}W_1+\mathbf{z}_{l-1, i}W_2$$

The results are then element-wise passed through a non-linear activation function called a \textit{gated activation},
$\tilde{\mathbf{y}}_l=\mathbf{y}_lW_{f,k}\odot sigmoid(\mathbf{y}_lW_{g,k})$.
Here $W_{f,k}$ and $W_{g,k}$ are learned parameters and $sigmoid$ is a non-linear activation function.
The output of these functions are added back to the input of the layer to get the final output of this layer. This technique is known as \textit{residual connections} which are used to address the problem of the gradient value becoming small~\cite{vanish} during gradient descent.

The output of the last dilated  layer is a $200\times8$ matrix, which is then compressed by a normal convolutional layer and a sigmoid activation function to produce a $200\times1$ result. This final layer uses a kernel size and filter size of one. Such an architecture compresses the input signal by $\frac{1}{8} = 12.5\%$, { and uses only $O(200L)$ computations during inference}.

\vskip 0.05in\noindent\textbf{Recurrent Neural Networks for adaptive classification}. 
We use a RNN in our architecture to gradually increase the confidence of our carrier sense function after each time step of 0.8~ms, which is smaller than the preamble size of all the considered LPWAN protocols. This allows us to adaptively select a window size for different protocols.

Unlike regular feedforward neural networks which contain no cycles, RNNs are a special kind of neural network topology {that ``memorize" states temporally. Specifically,} the output of an RNN layer is {not only passed to the next layer, but also} looped back and, {along with the next input,} provided as input to the RNN layer itself {at the next time unit}. 

If the RNN is unable to predict the presence of a signal at time $t$ with high confidence, it can still pass its output state to the next time step $t+1$ which can use this information to make a more informed prediction. After several time steps, the recurrent neural network accumulates enough information and eventually outputs the a high confidence value for protocols that require a longer preamble. On the other hand, for protocols that use a smaller preamble, the RNN can determine the existence of a signal after the first time period, $t$. Hence, we can achieve adaptive processing delays for different channel properties and protocols. 

In our implementation, the output of the RNN layer is passed through the \textit{ReLu} activation function. This is then passed to a final fully-connected layer with a $softmax$ function that produces the carrier sense output.

%% file: rate-2.tex
\section{Multiple rates using \name}

\begin{figure*}[ht]
        \begin{minipage}{0.32\textwidth}
        \includegraphics[width=1\textwidth]{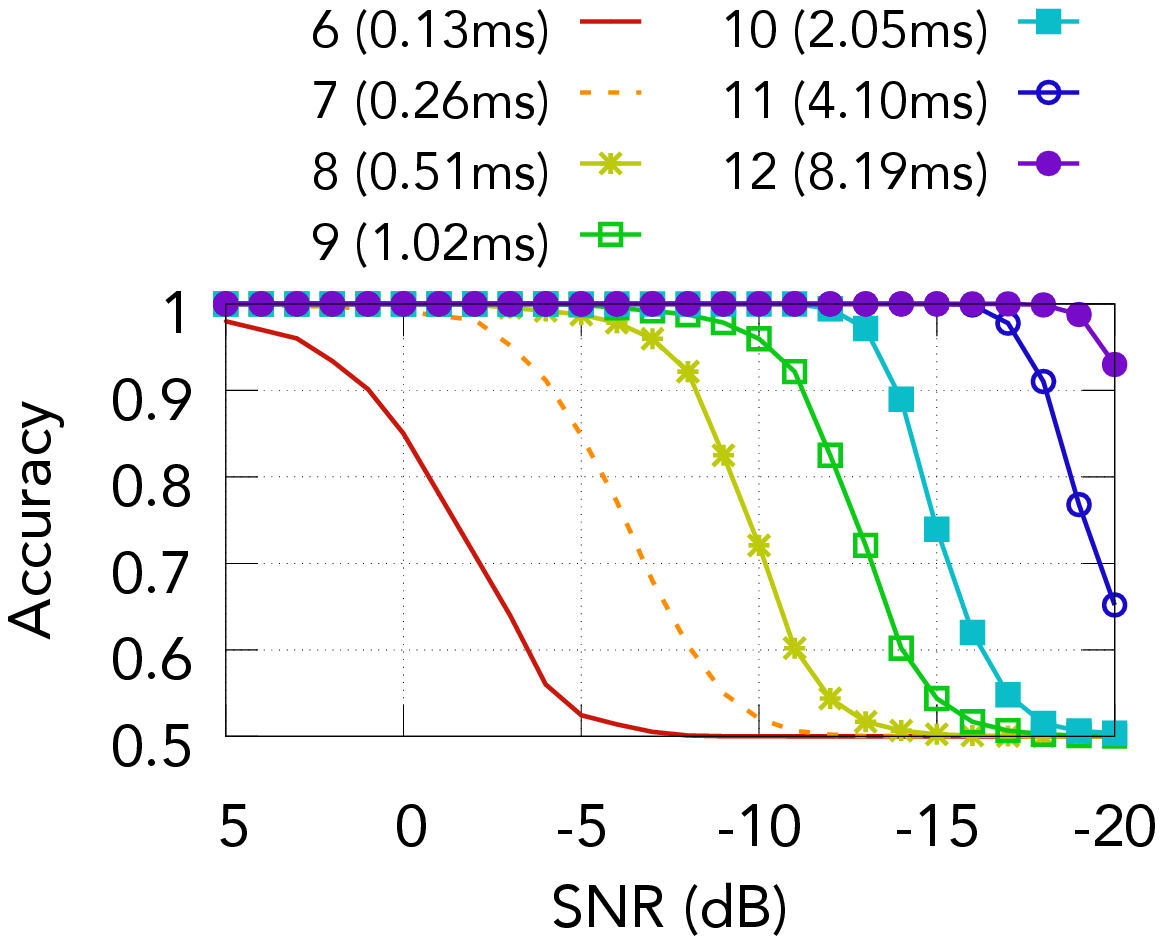}
        \caption{\textmd{\label{fig:baseline} LoRa baseline detection accuracy and the duration of a chirp.}}
        \end{minipage}
        \hspace{2ex}
        \begin{minipage}{0.65\textwidth}
        \begin{subfigure}{.49\textwidth}
                \includegraphics[width=\textwidth]{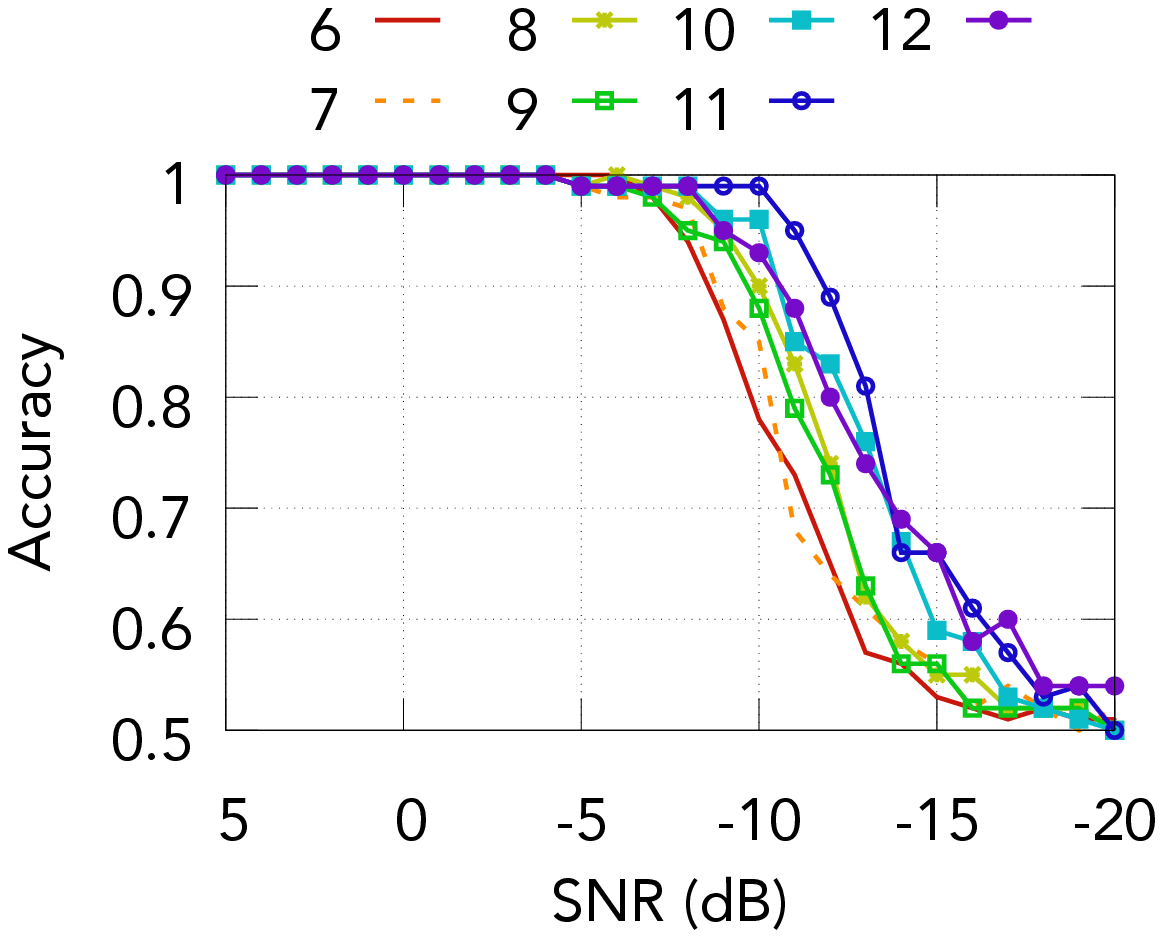}
                \caption{\textmd{8~ms carrier sense window}}
                \end{subfigure}
        \begin{subfigure}{.49\textwidth}
                \includegraphics[width=\textwidth]{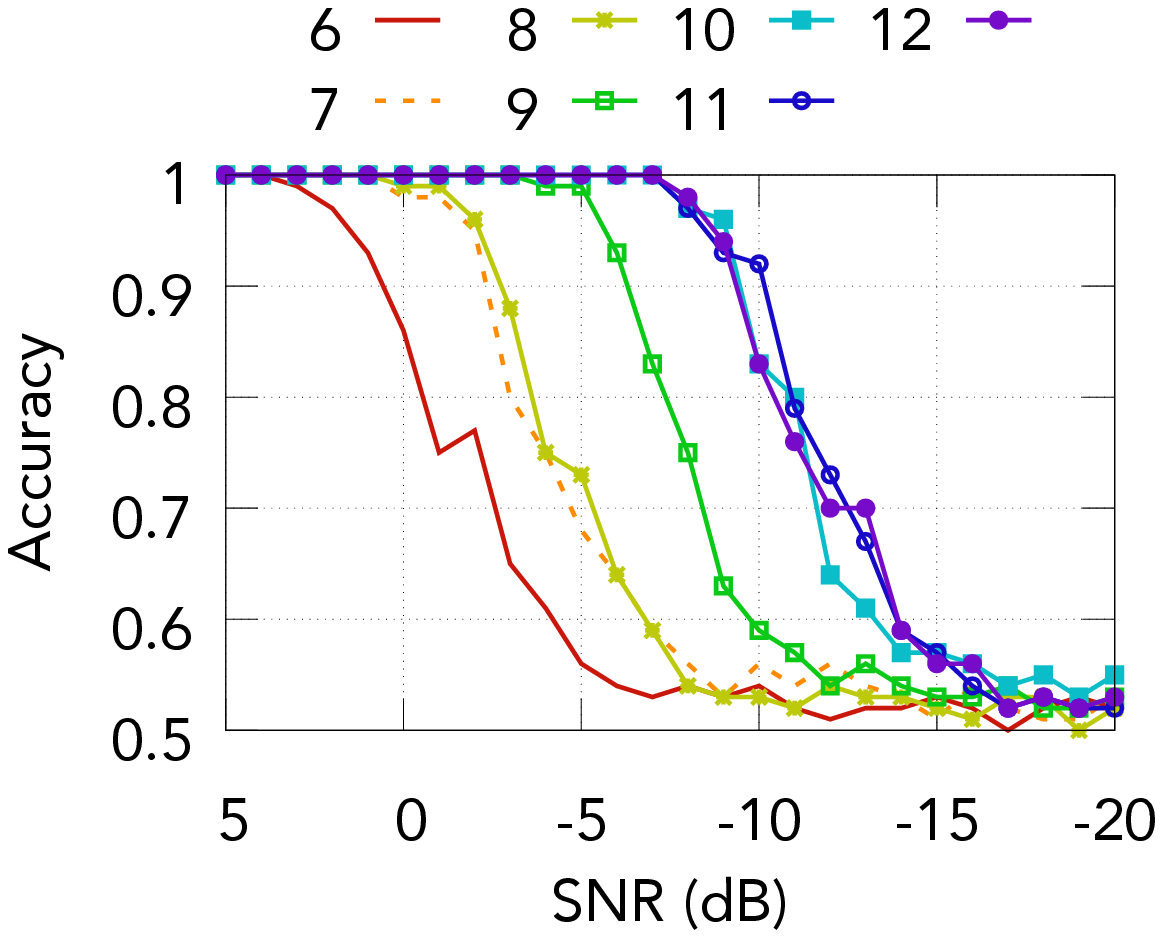}
                \caption{\textmd{1~ms carrier sense window}}
                \end{subfigure}
        \caption{\textmd{Carrier sense using spectrogram+CNN method.}}
		\label{fig:cs}
        \end{minipage}
        \vskip -0.15in
\end{figure*}

Beyond carrier sense, the ability to identify preambles of different configurations within a single protocol can enable LPWAN networks that can support multiple bit rates.
Specifically, while today's LPWAN protocols (e.g., LoRa) can support a large number of configurations. For instance LoRa supports a total of seven spreading factors (SF) and three bandwidths resulting in 21 different preambles. Requiring a single network to operate on all these 21 configurations requires the access point to decode all the corresponding 21 preambles, which is challenging in practice and hence today's network are configured to a single rate. 

An alternate solution that is used in Wi-Fi is to transmit the preamble at the lowest data rate (e.g., 6~Mbps) and use higher bit rates for the payload. While such a solution would work with large payload sizes (1500 bytes), sensor networks transmit tens of bytes in their payload and hence the overhead of the lower bit rate preamble can be prohibitive. To understand this  consider two scenarios. In the first scenario, the LPWAN device sends a packet to the access point by sending its preamble at the lowest data rate supported by LoRa, and its payload at the highest data rate. In the second scenario, the device sends its preamble and payload at the highest data rate. For LoRa's default preamble length and a 50 byte payload, this translates to 100.3~ms and 1.6~ms for preamble in the two scenarios and 12.5~ms respectively for the payload. This shows that the Wi-Fi approach of using the lowest bit rate preamble does not work in LPWAN networks.

We can however use \name{} to enable devices to transmit at different bit rates while using our deep learning framework to classify between different configurations at the receiver. This enables closer devices to transmit at a higher bit rate to AP and achieve longer ranges by supporting farther devices that transmit at a much lower bit rate.

To this end, instead of using two units at the final layer in the above architectures, we use 21 units to classify between different LoRa configurations. After learning on signals from the 21 different configurations, this can be used by the receiver to infer the LoRa configuration from the received signal and support multiple rates on the same network. Specifically, the access point sends periodic beacons  (one minute) at the lowest bit rate which each node uses to compute the RSSI. Using the AP's RSSI and channel reciprocity, each device picks the bit rate it can transmit its data by mapping to the sensitivity supported by each bit rate~\cite{loramod}. 

%% file: eval-2.tex
\section{Evaluation}
\label{sec:eval}

\begin{figure*}[ht]
        \begin{minipage}{0.23\textwidth}
        \includegraphics[width=1\textwidth]{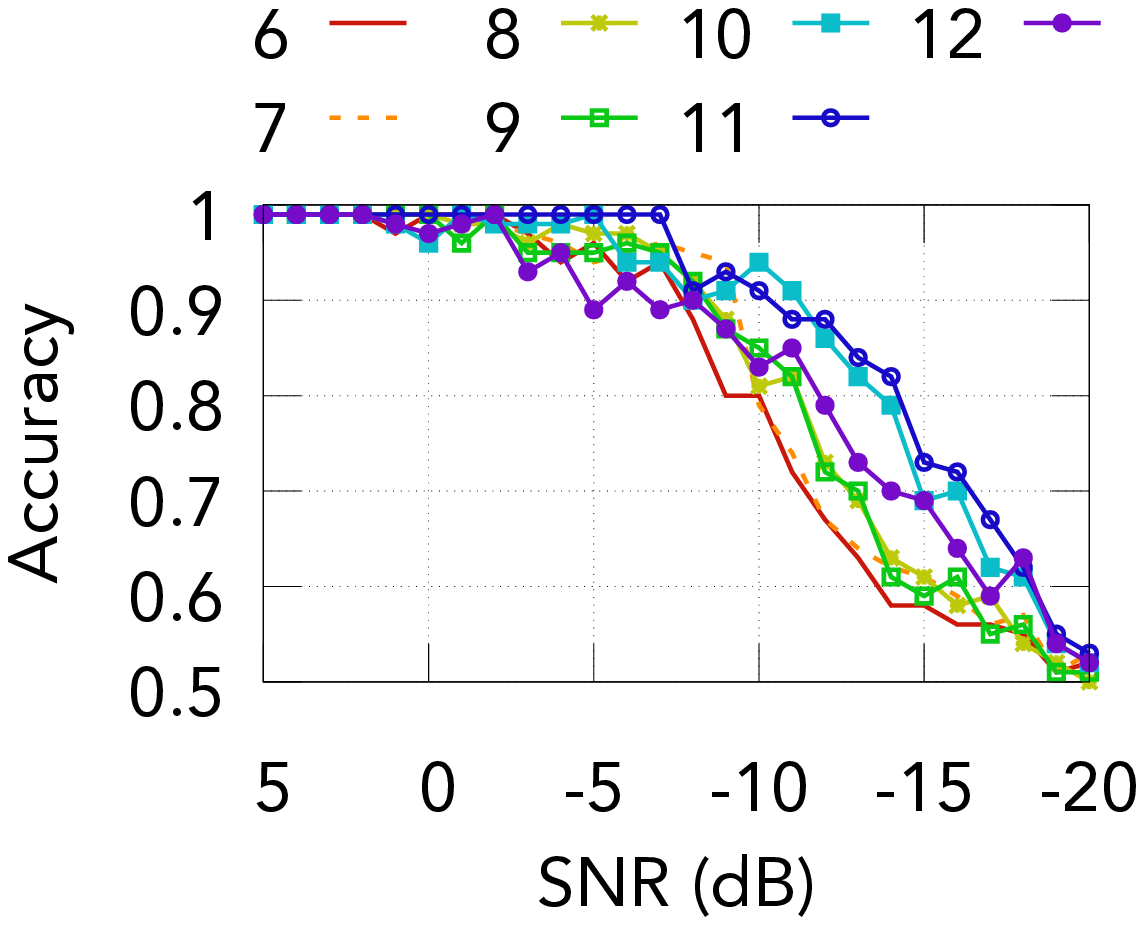}
        \caption{\textmd{Carrier sense in the presence of frequency offsets up to 250~kHz}}
        \label{fig:offset}
        \end{minipage}
        \hspace{1ex}
        \begin{minipage}{0.23\textwidth}
        \includegraphics[width=1\textwidth]{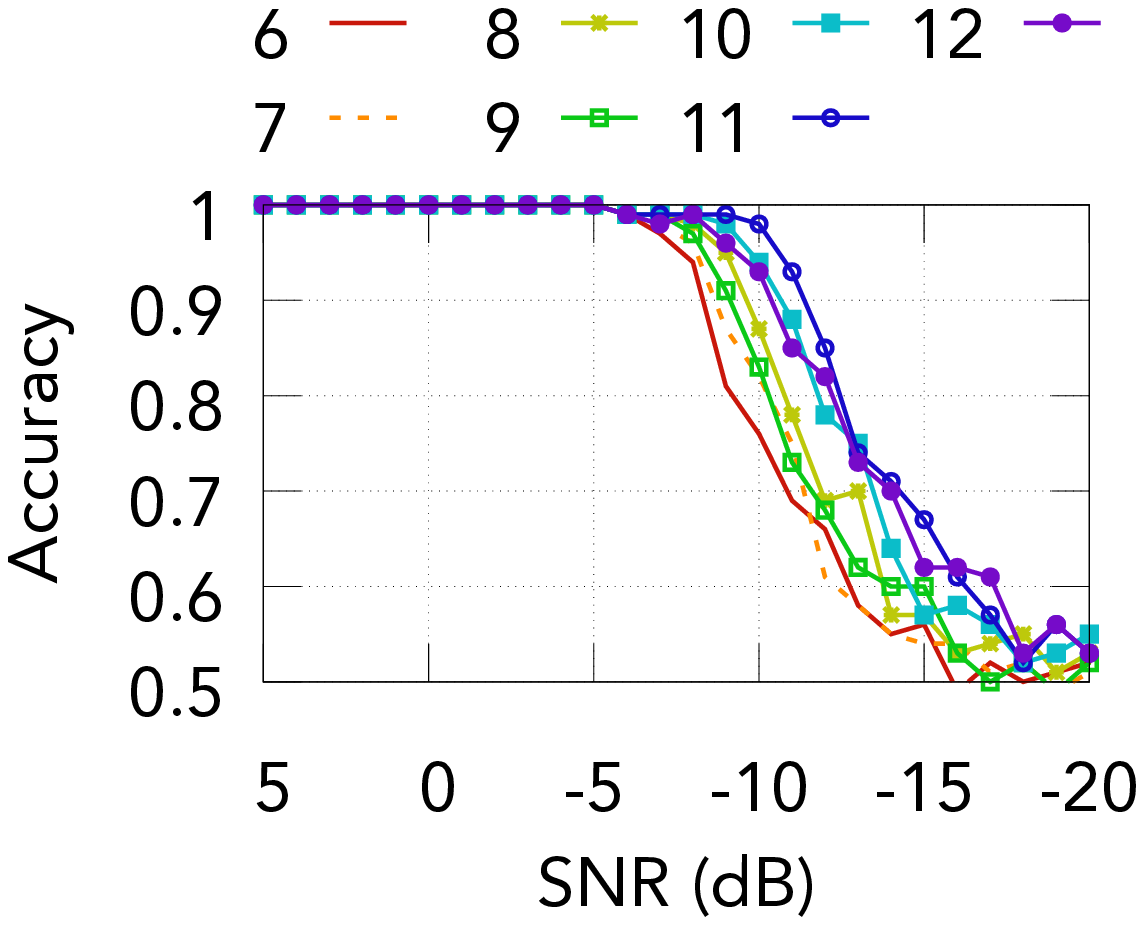}
        \caption{\textmd{Carrier sense in the presence of interference from NB-IoT* signals.}}
        \label{fig:interef}
        \end{minipage}
        \hspace{1ex}
        \begin{minipage}{0.50\textwidth}
        \vskip -0.3in
        \begin{subfigure}{.47\textwidth}
			\includegraphics[width=\textwidth]{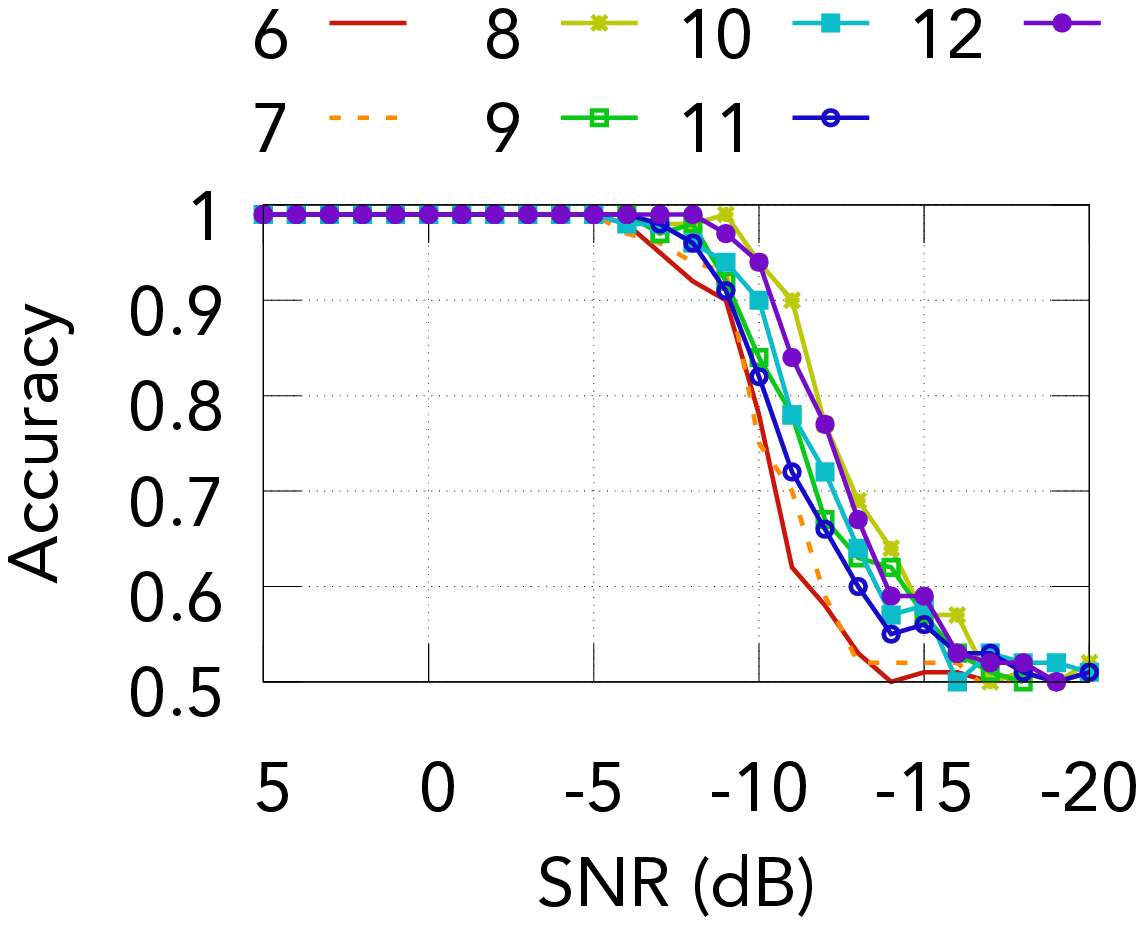}
            \caption{\textmd{500kHz + 500kHz signal}}
		\end{subfigure}
        \hspace{1ex}
		\begin{subfigure}{.47\textwidth}
			\includegraphics[width=\textwidth]{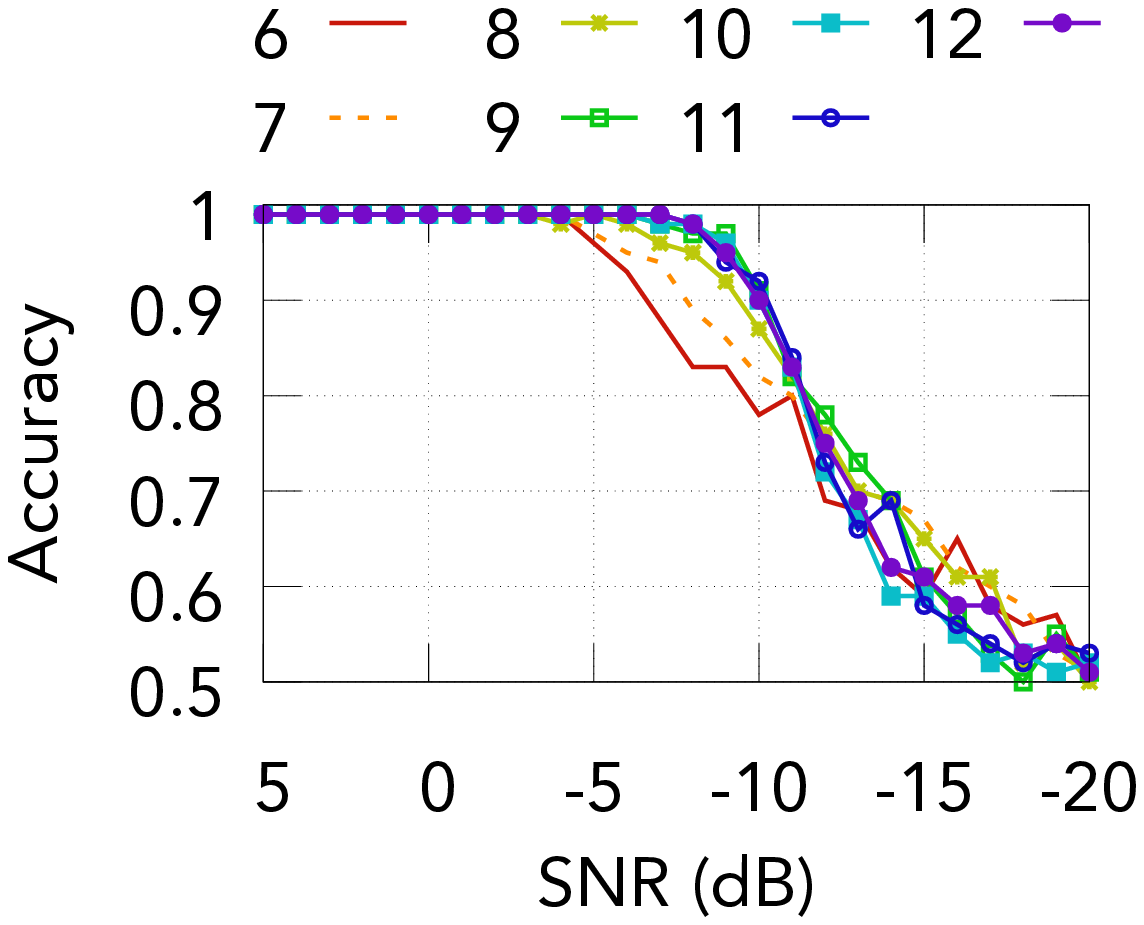}
            \caption{\textmd{500kHz + 250kHz signal}}
		\end{subfigure}
		\caption{\textmd{Carrier sense with concurrent transmissions}}
		\label{fig:concurrent}
        \end{minipage}
\end{figure*}

\input{training-2}

\subsection{Evaluating \name\ Carrier Sense}

\subsubsection{Carrier sense across LoRa configurations.} First, we start with the LoRa protocol and evaluate if \name\ can perform carrier sense across all the 21 different configurations. To do this, we  train our classifiers with examples from each of the 21 different LoRa configurations with artificially generated noise as described in~\xref{sec:exp}. The classifier includes training data of signals at an SNR of 5, 0, -5 and -10~dB. We include training data of a configuration at a given SNR only if it is designed to be detected at that sensitivity. For example, at an SNR of -10~dB we only include data from a spreading factor of 10--12. We do not include signals from lower spreading factors as that would be equivalent to training our model to recognize noise as signal. Empirically, we also find that training on a wide range of positive and negative SNRs and channel effects yields better accuracies compared to training on a single SNR or training on only higher SNRs.  

Fig.~\ref{fig:cs} shows the classification accuracy on the test data of our first approach using a spectrogram and CNN for different fixed carrier sense windows. The plots show the results for different spreading factor values as a function of SNR. Note that each spread factor combines the accuracies across the three bandwidth values of 125, 250 and 500~kHz.

To understand these results, we plot the baseline detection accuracies for a receiver that decodes each of the LoRa symbols in Fig.~\ref{fig:baseline}. The legend indicates the length of a chirp for each spreading factor, which is also the length of the baseline carrier sense window when the bandwidth is 500~kHz. An optimal LoRa decoder detects the signal by multiplying the signal by a downchirp of the corresponding bandwidth and spreading factor. Note that since different LoRa configurations have different downchirps with different spreading factors and bandwidth, they occupy different duration on the wireless medium. Further, the accuracy only depends on the spreading factor and not the bandwidth. Finally, we also note that these baseline measurements closely match the minimum SNR sensitivity as specified in LoRa datasheets~\cite{loramod}. 

Comparing Figs.~\ref{fig:baseline} and~\ref{fig:cs}(a) reveals that with a 8~ms carrier sense window, we achieve slightly higher accuracies  for spreading factors of 6 and 7 than the baseline detector. This is because the carrier sense window of the spectrogram is much longer than the baseline carrier sense window, which is only as long as a single chirp.
However, at spreading factors of 11 and 12, our accuracies are slightly worse than that achieved by the baseline detector using a downchirp. 
Finally, the lowest SNR at which \name\ can detect a signal reliably is at -11~dB for an LoRa transmission with a spreading factor of 11 and this can be done with a carrier sense accuracy of 95\%. Thus, \name\ can perform carrier sense below the noise floor.

\subsubsection{Carrier sense with frequency shifts} Next we test how robust our model is in the presence of transmissions at different center frequencies. This can happen in practice because a LPWAN transmitter could be transmitting in the second half of the receiver's 1~MHz bandwidth. Using the same model trained in the previous set of experiments, we generated a new set of test data where a 500~kHz LoRa signal was offset by a random frequency offset within the range [-250~kHz, 250~kHz]. This means that the LoRa signal can lie anywhere within the 1~MHz signal sampled at the receiver. 

Fig.~\ref{fig:offset} shows the carrier sense accuracies for these frequency shifted signals across all LoRa spreading factors. The plots show that there are no large differences in accuracies from the previous scenario with no frequency shifts. This is expected because convolutional neural networks that use a pooling operation are translation invariant and hence, they are able to accurately classify test inputs that have been offset in frequency as well as time from the training data. At an SNR of -10~dB the average decrease in accuracy as a result of the frequency offsets is 4\%. Note that our training data for the above model used maximum frequency offsets of 10~Hz but was able to carrier sense on frequency offsets up to 250~kHz. We can in principle increase our accuracy by growing our training set to include larger frequency offsets.

\begin{figure}[t]
\begin{subfigure}{.235\textwidth}
\includegraphics[width=\textwidth]{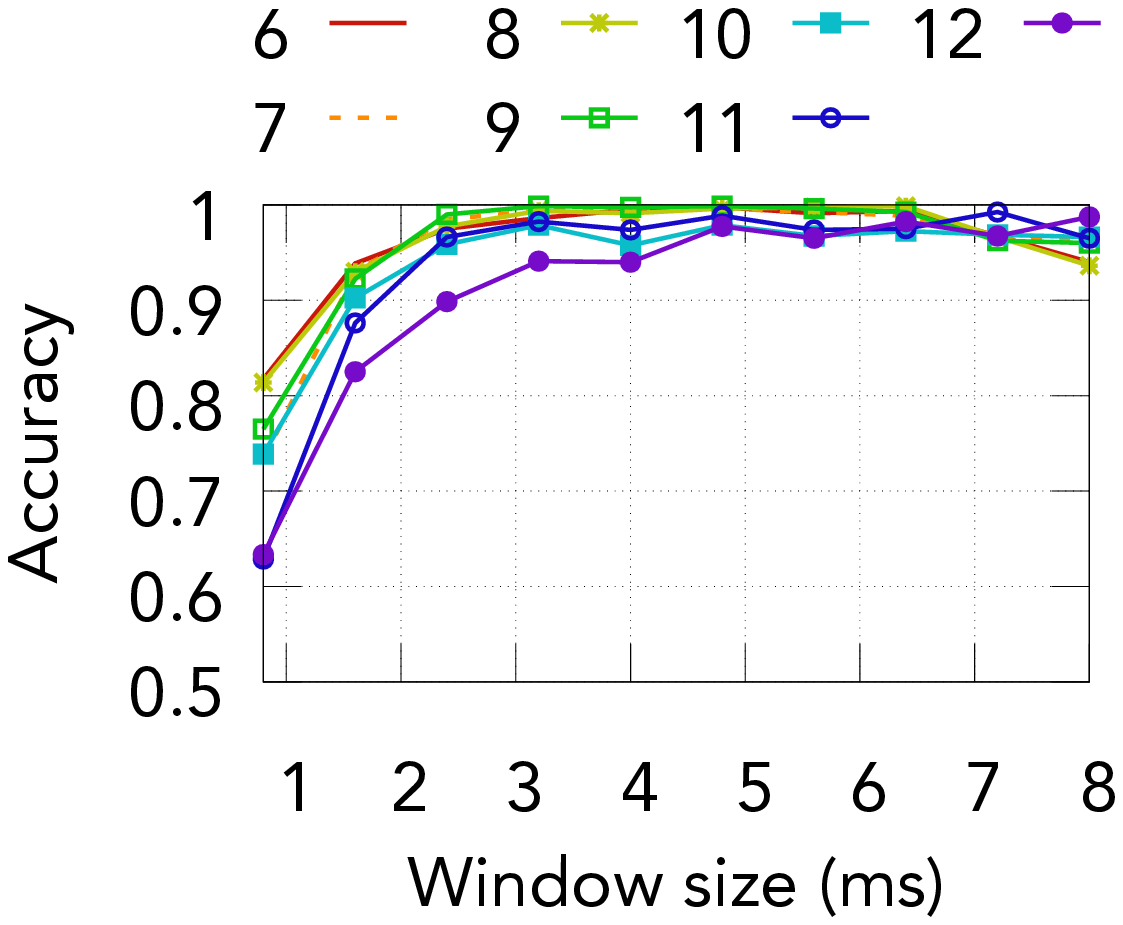}
\caption{\textmd{SNR from -10 to -5~dB}}
\end{subfigure}
\begin{subfigure}{.235\textwidth}
\includegraphics[width=\textwidth]{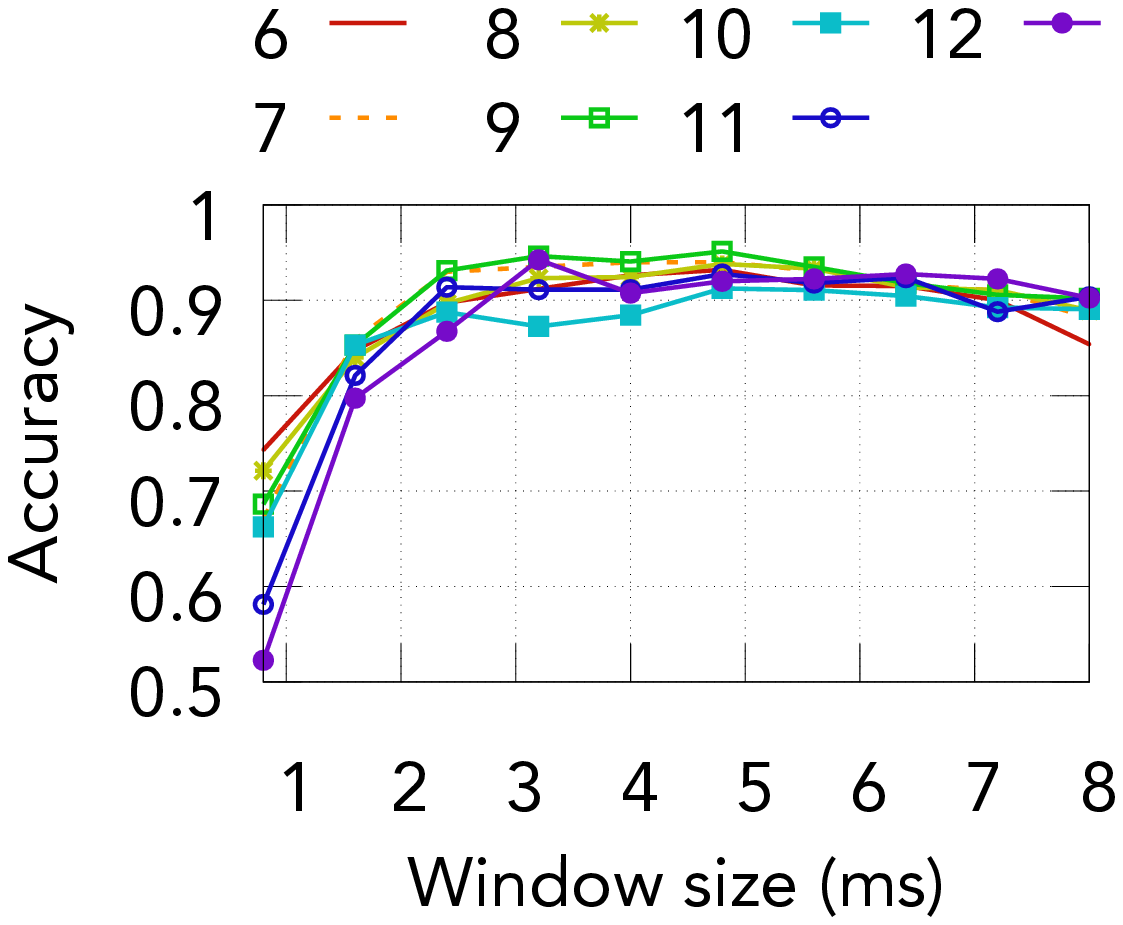}
\caption{\textmd{SNR from -15 to -10~dB}}
\end{subfigure}
\vspace{-0.15in}
\caption{ \textmd{Our dilated convolutions + RNN system has higher  accuracies when its window size increases.}}
\label{fig:rnn}
\vskip -0.15in
\end{figure}

\subsubsection{Carrier sense with concurrent transmissions.} Since the receiver is sampling with a bandwidth of 1~MHz, there could be multiple concurrent transmissions on the same band. We consider three different scenarios. First, we use two 500~kHz LoRa transmissions that are transmitting concurrently and are adjacent to each other in the frequency domain with similar SNRs. Second, we instead use two LoRa transmissions but now with 500~kHz and 250~kHz adjacent bands. Finally, we have a 500~kHz LoRa transmitter and NB-IoT* transmitter on the same set of frequency. The first two scenarios evaluate  carrier sense with two transmissions in adjacent bands within the received signal while the third scenario evaluates our ability to identify LoRa signals in the presence of interference from other protocols. 

To evaluate this we use the same model that was generated in the previous section on LoRa signals. We then test our model on test data in the above three scenarios. In Fig.~\ref{fig:concurrent}(a) and (b) we plot the accuracies in the presence of these concurrent transmissions at our 1~MHz bandwidth receiver. The plots show that there are no significant changes in the accuracies which again confirms the invariant nature of neural networks for spectral sensing. {Fig.~\ref{fig:interef} shows the carrier sense accuracy in the presence of an interfering NB-IoT* transmission. The plot shows that the carrier sense accuracies are high even in the presence of interference, across a range of SNRs. The average accuracy only reduces by 2\% at an SNR of -10~dB across the different LoRa configurations.}

\begin{figure*}[t!]
\begin{subfigure}{.19\textwidth}
\includegraphics[width=\textwidth]{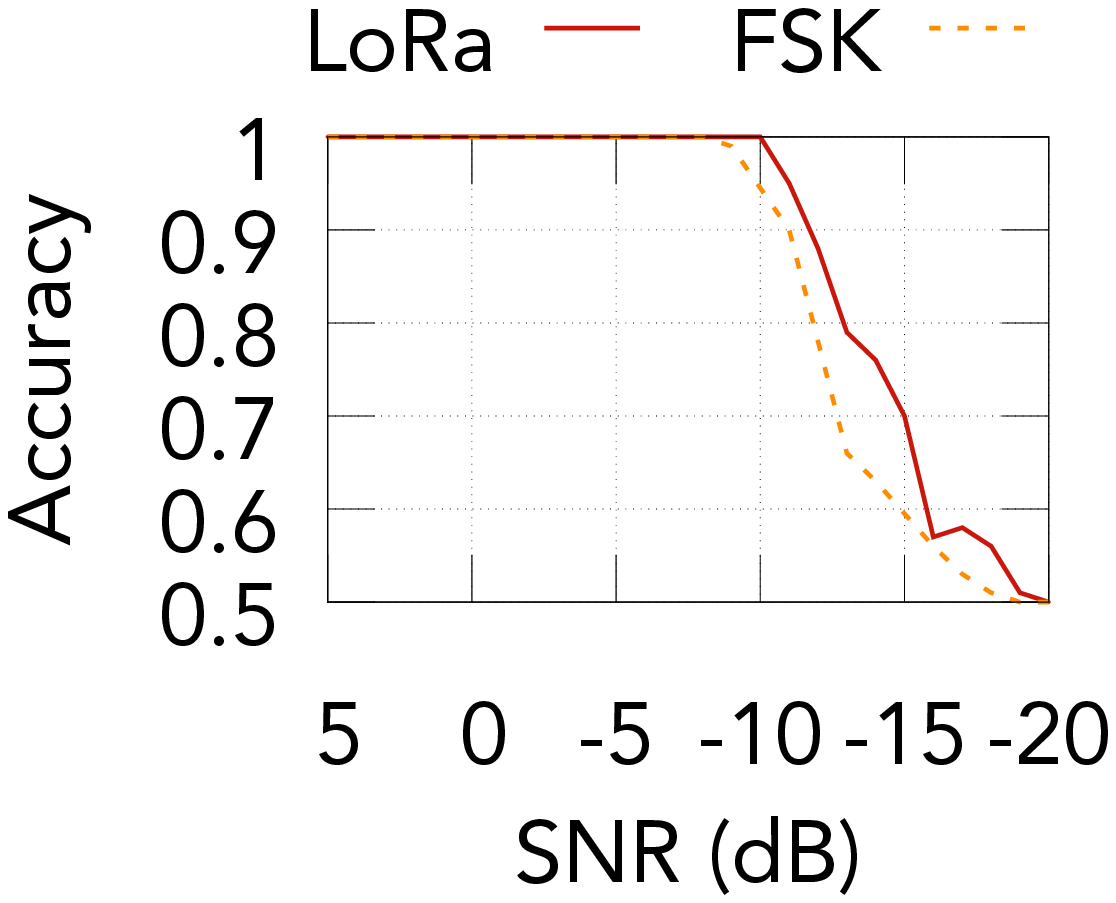}
\caption{\textmd{FSK}}
\end{subfigure}
\begin{subfigure}{.19\textwidth}
\includegraphics[width=\textwidth]{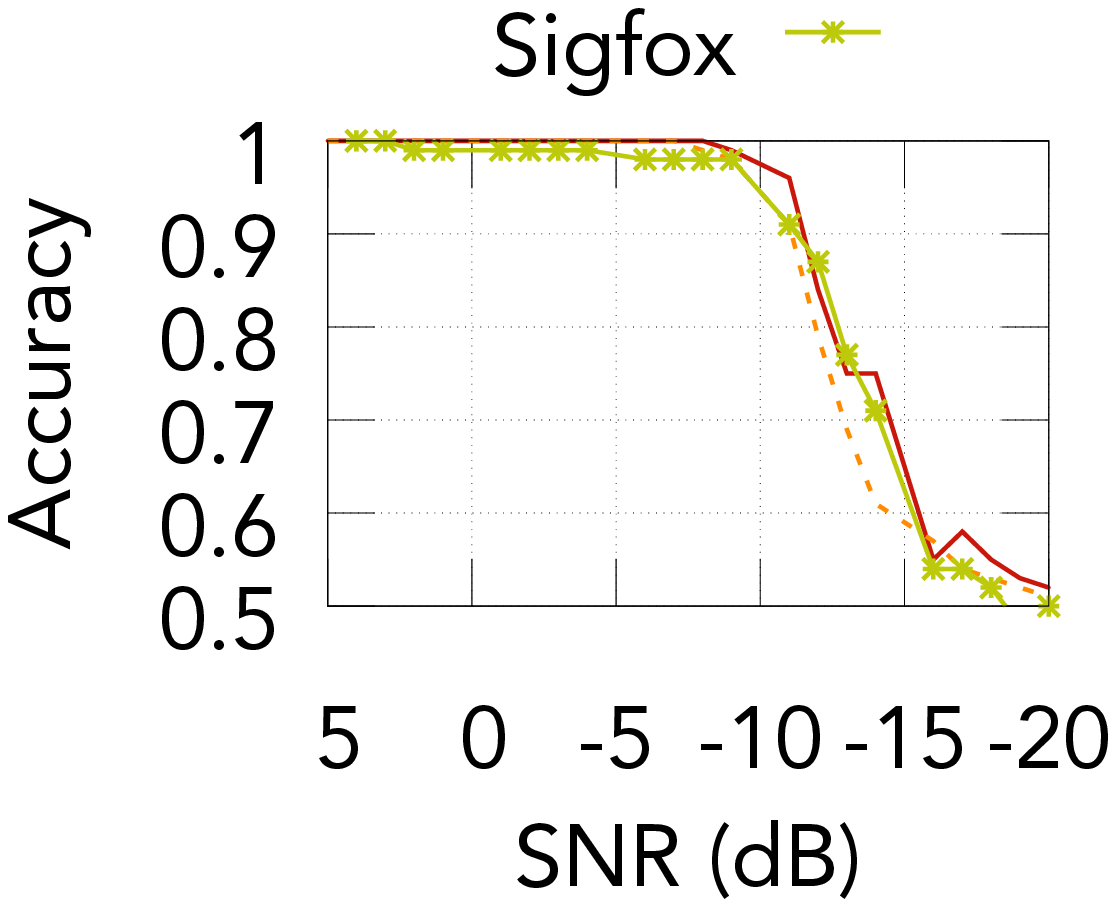}
\caption{\textmd{Sigfox}}
\end{subfigure}
\begin{subfigure}{.19\textwidth}
\includegraphics[width=\textwidth]{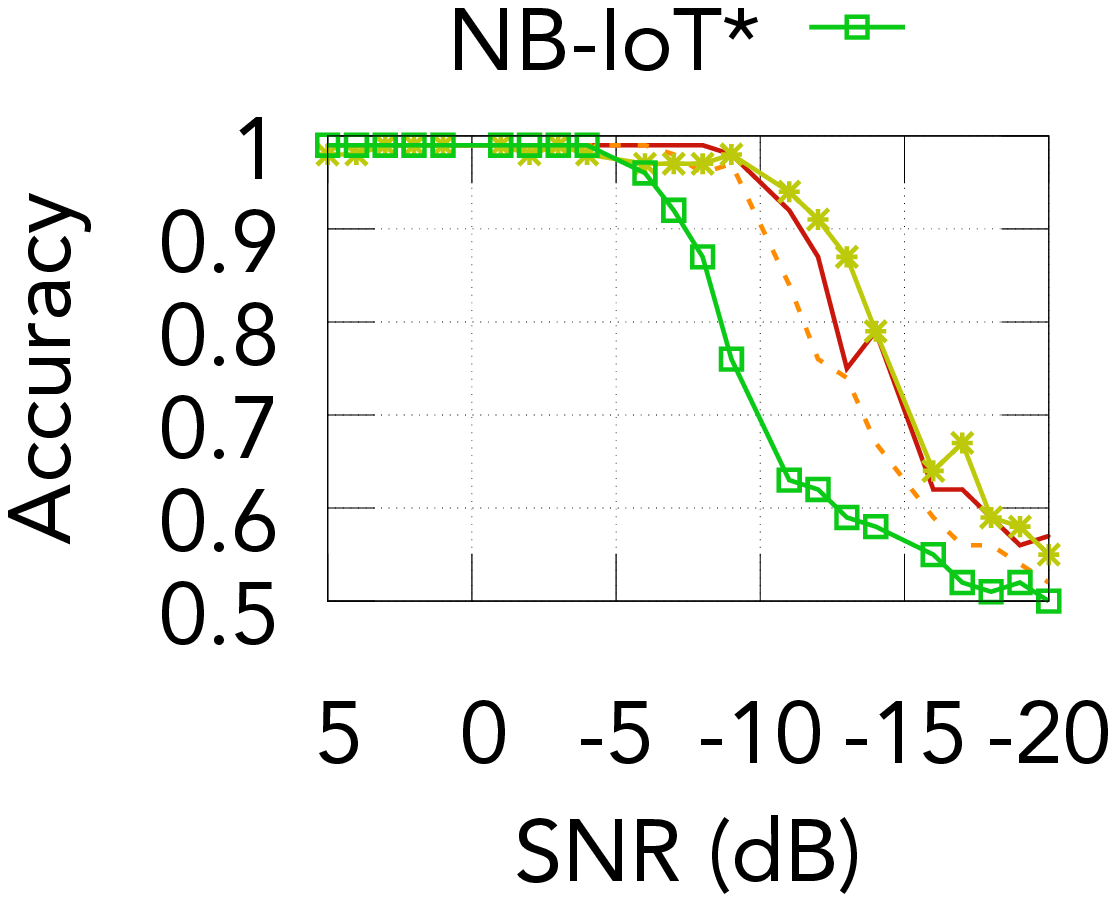}
\caption{\textmd{NB-IoT*}}
\end{subfigure}
\begin{subfigure}{.19\textwidth}
\includegraphics[width=\textwidth]{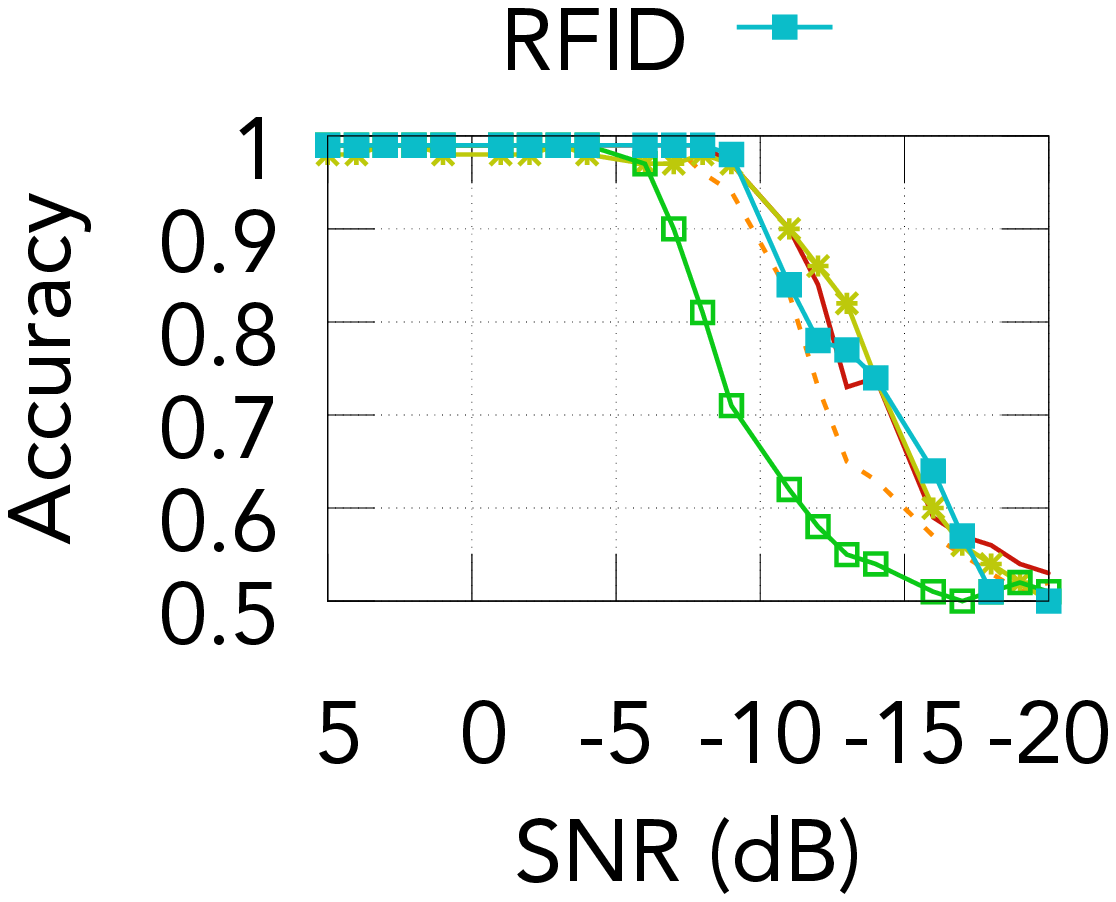}
\caption{\textmd{RFID}}
\end{subfigure}
\begin{subfigure}{.19\textwidth}
\includegraphics[width=\textwidth]{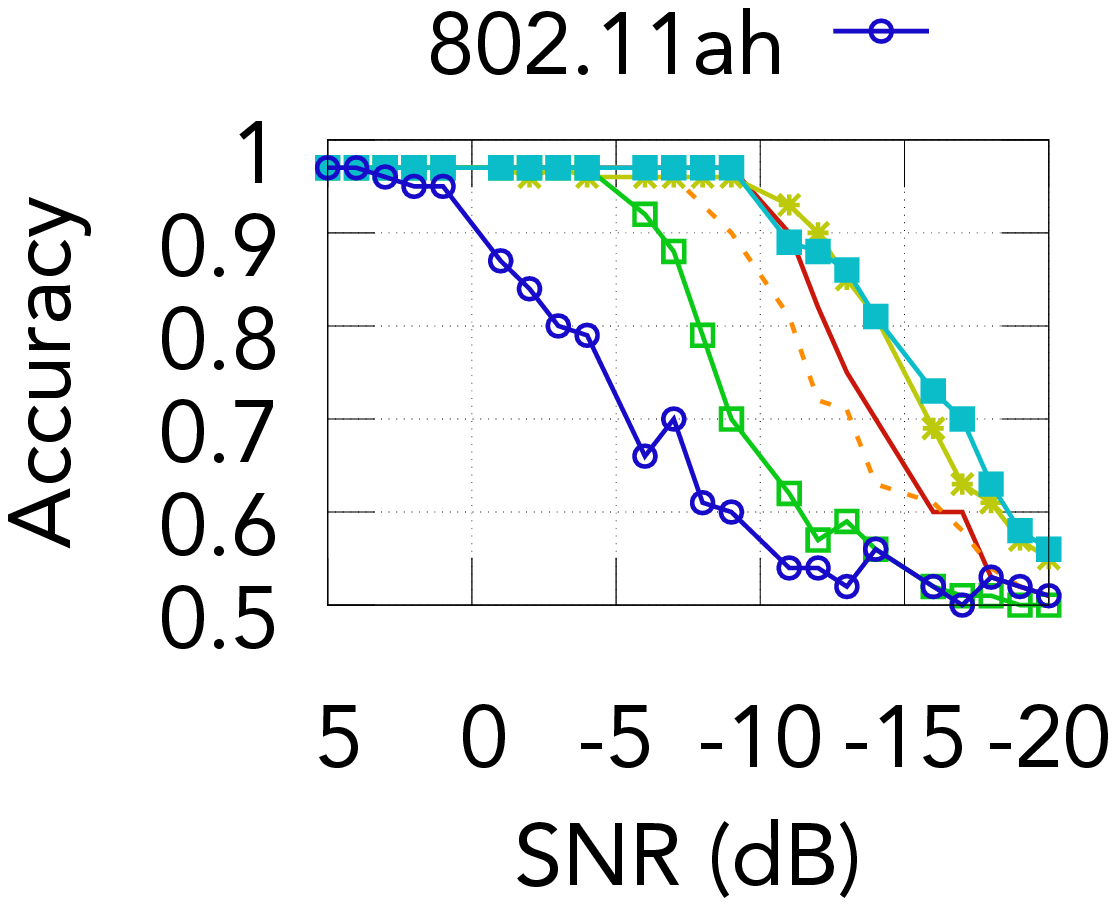}
\caption{\textmd{802.11ah}}
\end{subfigure}
\vskip -0.15in
\caption{\textmd{\name{} generalizes across protocols.}}
\vskip -0.2in
\label{fig:gen}
\end{figure*}

\subsubsection{Variable carrier sense window.} 
{To evaluate the adaptive window size capability of our dilated CNN and RNN approach, we train a LoRa carrier sense classifier with the same data as the above model. We then test our system's accuracy at different window sizes. Figure~\ref{fig:rnn} shows the accuracies for different spreading factors under two SNR ranges namely -10 to -5~dB and -15 to -10~dB. 

The plots show that our RNN classifier achieves higher accuracies with larger window sizes. The major accuracy gains occur when the window size increases from 0.8~ms to 3.2~ms, after this point there are diminishing returns on the accuracy gains. The main benefit of this approach is that we can adaptively test the performance of our system on different window sizes when testing our classifier. Depending on the requirements of our carrier sense application, we can use a small window size if our main concern is latency, or a larger window size if we care more about accuracy, without changing the topology and any parameters in the DNN architecture. This is unlike the spectrogram method which requires committing to a window size before preprocessing the data and finalizing the DNN architecture and its parameters, and may result in a either long latencies with a large window size or low accuracies due to a small window size.

We also note that compared to the spectrogram approach,  the carrier sense performance of the RNN approach is much better at low SNRs below -10~dB. Specifically, it can achieve accuracies of 88\% while the spectrogram can achieve average accuracies of only 61\% in these SNR ranges. We believe this is because the RNN architecture uses the phase information which the spectrogram approach discards. This allows for our RNN architecture to learn more information and achieve better accuracies at lower SNR regimes.
}

\subsubsection{Generalization and forward-compatibility.} Finally, we evaluate how well our carrier sense architecture generalizes when we wish to support carrier sense for multiple protocols at the same time. To do this we first trained our deep learning system on LoRa SF12 signals and noise samples. We then added additional signals from LoRa FSK to the train set to obtain a new set of carrier sense accuracies for each protocol across SNRs. We repeated this process by incrementally adding a new protocol to the train and test sets. We added Sigfox, NB-IoT*, RFID and 802.11ah traces in that order. We emphasize that the same deep learning architecture, with the fixed number of weights and layers, was used when training each collection of protocols.

Fig.~\ref{fig:gen} shows the accuracies for each collection of protocols across SNRs. The plots shows that as the number of protocols added to the classifier increase, the detection accuracies for individual protocols generally stay the same. We find that we can detect certain protocols at sensitivities that were better than they were designed for. In particular, we can detect RFID signals down to -9~dB. 
We also note that 802.11ah is designed to operate above 0~dB. Further, the accuracies at positive SNRs across all the protocols, decrease to 97\% when we add 802.11ah. This is because unlike the RNN approach which uses phase, the spectrogram of the OFDM signal occupies the entire 1~MHz band and looks like high noise in terms of its spectral properties. A general solution  would be to either use the RNN architecture or increase the receiver sampling rate to be larger than the largest bandwidth of signals in our training set. This way, the classifier can recognize the signal as a band in a larger window. 

%% file: training-2.tex
\subsection{Experimental Methodology} 
\label{sec:exp}
{\it Training dataset.} We capture over-the-air transmissions from a LoRa transmitter that supports 21 different LoRa configurations, a LoRa FSK transmitter as well as Sigfox, NB-IoT* and 802.11ah transmitters in a {\it single location}. Using on-air transmissions ensures that the data captures various practical considerations such as sampling and frequency offsets. We then artificially simulate and introduce different wireless channel effects and noise to the training data set. 

Specifically, LoRa signals are transmitted using a Semtech SX1276RF1KAS~\cite{lorachip} and MSP430FR5969 LaunchPad Development Kit~\cite{loramcu}. We collected 1000 LoRa packets with a randomized payload for each of LoRa's 21 physical layer configurations. We repeated the same process to collect FSK modulated LoRa signals. Each signal has a four byte payload. Similarly, Sigfox packets were transmitted using the Wisol WSSFM10R2 Breakout Board~\cite{sigfoxchip} at 100~bps. For 802.11ah, we were not able to find a commodity 802.11ah chip to send arbitrary packets, so we generated the 802.11ah signals in software with the WLAN system toolbox~\cite{wlan} and transmitted them over the air using a separate unsynchronized USRP. Our signals had a bandwidth of 1~MHz and used an MCS of 0 and BPSK. Finally, while NB-IoT uses cellular bands, we transmit them on the 915~MHz ISM bands using the LTE system toolbox~\cite{lte} to evaluate realistic future LPWAN protocols in the ISM band.  All of our signals are captured by a USRP on a FLEX900 daughterboard with a sampling rate of 1MS/s. 

On this over-the-air data, we apply additional distortions to the signals so that we enrich our dataset to generalize to a wider array of channel conditions. We apply frequency offsets of up to 10~Hz, phase offsets and Doppler shifts. We also introduce Rician and Rayleigh multipath fading.
We use 90\% of the dataset for training and 10\% for validation. 

\vskip 0.05in\noindent{\it Preventing overfitting.}
We apply batch normalization~\cite{batch} which is a regularization technique that normalizes the outputs of each neural network layer. 
This technique is defined as:
$BN(x_{batch})=\frac{\gamma(x_{batch}-\mu)}{\sigma^2}+\beta
$,  where $\mu$ and $\sigma$ are the mean and variance of $x_{batch}$ to normalize $x_{batch}$, and $\gamma$ and $\beta$ are the scaling factors that are learned during training to transform the batch to match a desired distribution. 
We also use Lasso regularization to add a penalty term to our loss function in the form of an L1 regularization to prevent overfitting. 

\vskip 0.05in\noindent{\it Training complexity.} Training the weights for our first architecture takes less than an hour. Training for the second architecture is however  a time-consuming task that takes tens of hours even on a GPU. To accelerate this, we split the neural network into two parts before the RNN layer. For the first part, we connect its output directly to a dense layer to produce the final result. This new neural network is first trained using our training set. This takes several hours using a NVIDIA GeForce GTX 1060 GPU~\cite{gpu}. After that, we transfer the learned weights of this neural network into the original full neural network and train it again. For the RNN, we use the truncated back-propagation through-time algorithm~\cite{time} which takes about one hour on the GPU. 

\vskip 0.05in\noindent{\it Test dataset.} To ensure generalization, we do not use our training data for our testing purposes. Our test data is collected across eleven locations to span the whole operational SNRs for each of the tested protocols. This ensures that we are evaluating generalization across locations, over the air and different RF environments. We use all of our hardware including LoRa, Sigfox, NB-IoT* and 802.11ah. We also test with various configurations for LoRa including the 21 settings as well as FSK modulation. In a majority of our tested locations, there were RFID readers deployed and operational at the same time as our experiments. 

%% file: rate-eval-2.tex
\subsection{Evaluating Multi-rate LPWANs}
\label{sec:rateeval}

\begin{figure}[t]
\vskip -0.1in
\begin{subfigure}{.25\textwidth}
\includegraphics[width=\textwidth]{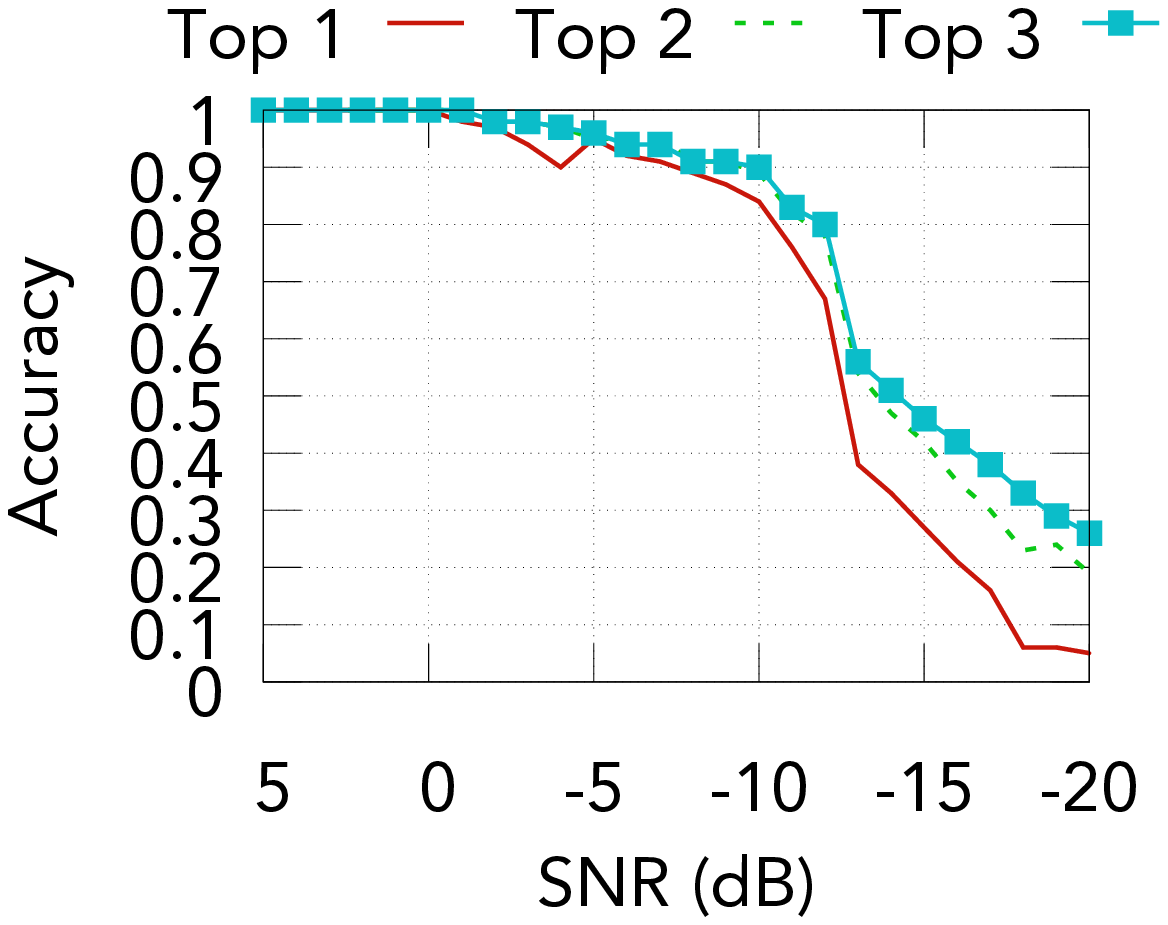}
\end{subfigure}
\begin{subfigure}{.08\textwidth}
\includegraphics[width=\textwidth]{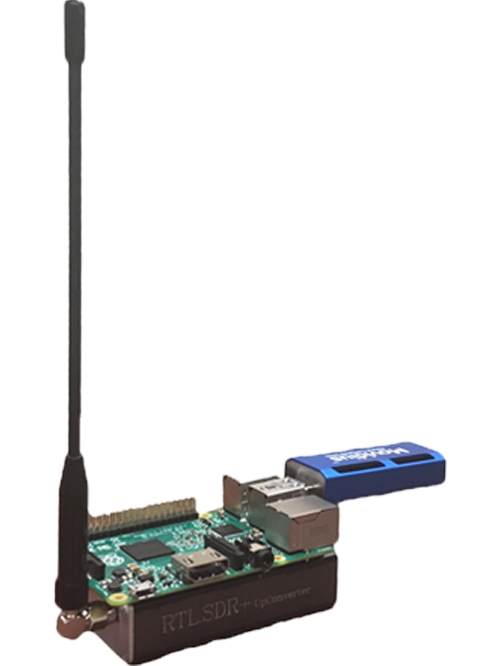}
\end{subfigure}
\vskip -0.15in
\caption{\textmd{(Left) Accuracy when classifying between different LoRa configurations. (Right) Our hardware prototype.}}
\label{fig:config}
\vskip -0.15in 
\end{figure}

To enable multi-rate networks, our \name{} hardware uses the same deep learning architecture as our carrier sense system except the number of units in the last layer to differentiate between all 21 different LoRa  configurations. 

We trained our deep learning system with signals across all 21 LoRa configurations at SNRs of 5, 0, -5 and -10~dB. At each SNR point, we trained and tested configurations that were detectable at that sensitivity. We then measure the classification accuracies across the  different configurations using our test data. As with the training data, at each SNR, the test dataset only considers the LoRa configurations that can be decoded at that SNR. For example, since SF 6 does not work below -10~dB even in the ideal scenario, we do not use it for testing for SNRs and the corresponding locations below -10~dB.

Fig.~\ref{fig:config} shows the accuracy of our LoRa configuration classifier from 5~dB to -20~dB with our spectrogram+CNN approach using a buffer size of 8~ms. {The plots show that \name\ can classify between the 21 LoRa configurations, with an average accuracy of 95\% for at SNRs from [-10,5]~dB. We note that a random guess between 21 classes  results in a classification accuracy of 4.7\%. Further, the accuracy that the desired configuration is within the top two or three predictions made by the classifier is higher at lower SNRs. At an SNR of -10~dB, a single class prediction yields a 84\%, however the probability that the correct class lies within the top two and three predictions is 89\% and 90\% respectively.}

To evaluate the benefits of a multi-rate LoRa network, we compare between two different scenarios. 
\squishlist
\item {\it Fixed bit rate.} All the devices in the LoRa network are set to a pre-determined bit rate of 9.38~kbps similar to prior work~\cite{lora-sigcomm17}, which is the existing approach for LoRa. 
\item {\it Multi-bit rate.} {Each of the devices in the network use a different bit rate by mapping RSSI values to the lowest spreading factor possible at that sensitivity~\cite{loramod} with a 500~kHZ bandwidth. The receiver then uses DeepSense to classify between the configurations and then decode the signals.}
\squishend

To evaluate the above two approaches, we set a LoRa transmitter in a fixed location and change the DeepSense receiver location across 30 different locations on our campus. In each of the locations, we measure the bit rates used by the transmitter using the above two approaches.

\begin{figure}[t]
\vskip -0.05in
\includegraphics[width=.43\textwidth]{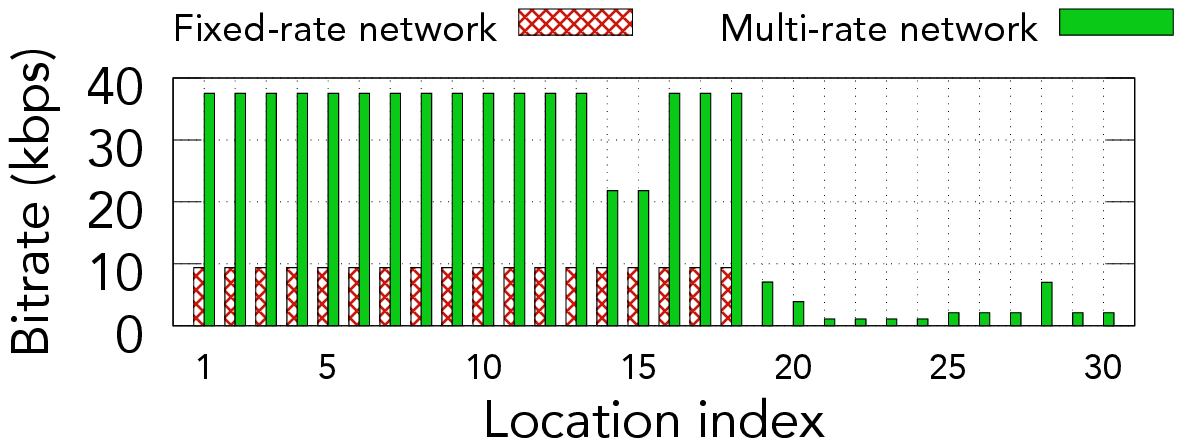}
\vskip -0.15in
\caption{\textmd{Throughput gains in a multi-rate LoRa network.}}
\label{fig:mapresult}
\vspace{-0.15in}
\end{figure}

Fig.~\ref{fig:mapresult} shows the selected bitrates in the multi-rate and fixed-rate network scenarios. The plot shows that at nearby locations, DeepSense can achieve a bit rate of 37.5~kbps which is 4x more than that achieved by the fixed bit rate solution. Further, as expected the number of locations that can connect to the network increase by a factor of 1.7x. This is expected because with rate adaptation the devices can operate across the whole range of LoRa bit rates from 290~bps to 37.5~kbps.

%% file: complexity-1.tex
\subsection{Complexity and Power Analysis}
\label{sec:comp}

{\it Off-the-shelf prototype.} We build a hardware platform using off-the-shelf hardware in Fig.~\ref{fig:config} that allows us to perform carrier sense in realtime. Our platform consists of a Raspberry Pi 3 which is connected via USB to a SDR~\cite{yoosoo} and the Intel Movidius machine learning accelerator that can execute inferences at 100~GFLOPS~\cite{movidius}. 
The SDR provides us with complex samples which are then streamed to our machine learning classifier which is implemented with the Keras framework~\cite{keras} using a TensorFlow backend~\cite{tensorflow}. }




{\it Power analysis.} A drawback of using Movidius is that it does not support efficient duty cycling, and runs inferences continuously even when a node does not need to transmit information. Additionally it is only configured to run at 100~GFLOPS at 1.2~W. However, our architecture requires two orders of magnitude less FLOPS to operate.

So we instead provide an estimate of the power consumption required to perform carrier sense on an ASIC. To do this, we first run TensorFlow's profiler to provide the number of floating point operations required to make a single inference. Given the number of inferences per second, assuming continuous operation, we then compute the FLOPS.
Then we follow the estimation method used in a recent implementation of an neural network ASIC accelerator~\cite{DNN_ASIC} to estimate the power consumption of our architectures. 
Specifically, we calculate the number of  arithmetic and memory access operations, and multiply each operation by the corresponding amortized energy consumption when using a 45nm CMOS process. 
After this we add the standby energy consumed by the ASIC's clock network, registers, combinatorial circuits and memory.
Table~\ref{tab:complexity} lists complexity and power consumption estimates of our models.
Note that state of the art 28~nm deep learning ASICs can consume less than a milliwatt~\cite{power1,power2,power3}.  These numbers are well within the power budget of LPWAN transceiver chips which typically consume 30-50~mW~\cite{power}.

\begin{table}[t]
\centering
\footnotesize

\begin{tabular}{|l|l|l|}
\hline
Model                           & \shortstack[l]{Spectrogram\\ +CNN} & \shortstack[l]{Dilated CNN\\ +RNN} \\
\hline
\# of parameters                     & 11,394          & 15,321           \\
\hline
FLOP per inference              & 2,789,504         & 679,441         \\
\hline
FLOPS                           & 348M            & 849M            \\
\hline
ASIC power estimate &9.95 mW & 11.08~mW\\
\hline 
\end{tabular}

\caption{\textmd{Complexity and Power Consumption}}
\vskip -0.2in 
\label{tab:complexity}
\end{table}

%% file: related-3.tex
\section{Related Work}

\vskip 0.05in\noindent{\bf Deep learning based communication.}
Over the past year, deep learning has attracted significant interest from the wireless theory community for its use in enabling wireless communication.~\cite{DARPA-program,polar,iclrcomm,hamming,ldpc} have used neural networks to learn various coding techniques including polar codes~\cite{polar}, random codes, convolutional codes~\cite{viterbi3,viterbi2,iclrcomm}, turbo codes, hamming~\cite{hamming} and LDPC codes~\cite{ldpc}. These approaches are able to learn the coding structure of signals and decode the bits when they are above the noise floor, i.e., SNR $\ge$ 0~dB. 

Deep learning has also been used for demodulation~\cite{survey1,survey2}.~\cite{air} shows decoding of DQPSK signals when the SNR is above 0~dB. ~\cite{mimo} extends these techniques to MIMO systems to decode bits from spatially multiplexed signals.

In contrast to prior work, our method can classify between various LPWAN protocols that use a variety of modulation and coding techniques, {\it below the noise floor}. Further, we show for the first time that one can  classify chirp spread spectrum signals at SNRs as low as -10~dB using deep learning.

\vskip 0.05in\noindent{\bf Cognitive radios.} The ability to identify the radio type and spectral occupancy has been a key research thread in the cognitive radio literature~\cite{grinspector}. Systems such as RFDump~\cite{rfdump} use energy detection to extract timing information about the packets and classify between wireless protocols such as Wi-Fi, Bluetooth and ZigBee. However, energy detection can be used to detect the presence of signals only when they are significantly above the noise floor.~\cite{whitespaces} uses correlation with a known preamble to detect radio types. Jello~\cite{jello} achieves spectrum occupancy sensing using edge detection on the power spectral density of the received signal. 

DoF~\cite{dof} uses the cyclo-stationary properties of communication systems~\cite{cyclo1,cyclo2} such as  Wi-Fi, Bluetooth, Zigbee, cordless phones as well as analog signals from microwave ovens to build unique signatures for each signal type and classify them using an SVM. DoF can classify between the above five 2.4~GHz wireless technologies in the presence of interfering signals and at SNRs at or greater than 0~dB.

In contrast to these approaches, \name{} is targeted for low-power wide-area protocols which are unique in that they require real-time operation, low-power consumption and can operate significantly under the noise floor. We perform carrier sense in the presence of various LPWAN technologies including LoRa, Sigfox and  NB-IoT* using deep learning.

%% file: conc-1.tex
\section{Discussion and Conclusion}
We present \name, the first carrier sense scheme that enables random access and coexistence for LPWANs. Here we outline two research opportunities to improve our design.

\vskip 0.05in\noindent{\it 1) Hidden terminals.} As is true with any carrier sense based system (e.g., Wi-Fi), we need to address hidden terminals and the resulting collisions. Recent work on Choir~\cite{lora-sigcomm17} can enable decoding of LoRa collisions in LPWANs, which can be useful in the presence of hidden terminals. Designing  collision decoding schemes that use deep learning across LPWAN protocols,  would be an interesting research direction.

\vskip 0.05in\noindent{\it 2) Exposed terminals and enabling concurrent transmissions.} While we motivate carrier sense to avoid concurrent transmissions, in some scenarios based on the signal strengths and the protocols involved, multiple transmissions should occur at the same time to increase the network throughput. Such designs have been explored for carrier sense and exposed terminals in Wi-Fi systems~\cite{harinsdi,romitcollision}. Since our design can not only perform carrier sense but also identify the specific protocol and configuration of the received signal, one can develop similar techniques to enable concurrent transmissions, which are dependent on the protocols in  the signal. 